\documentclass[12pt,preprint]{aastex}

\usepackage{emulateapj5}
\shortauthors{White, Helfand, Becker, et al.}
\shorttitle{I-Band-Selected Radio Quasars}

\begin{document}
\title{An I-Band-Selected Sample of Radio-Emitting Quasars: Evidence
for a Large Population of Red Quasars\altaffilmark{1}}
\author{
Richard~L.~White\altaffilmark{2},
David~J.~Helfand\altaffilmark{3},
Robert~H.~Becker\altaffilmark{4,5},
Michael~D.~Gregg\altaffilmark{4,5},
Marc~Postman\altaffilmark{2},
Tod~R.~Lauer\altaffilmark{6},
\&
William~Oegerle\altaffilmark{7}}
\email{rlw@stsci.edu}

\altaffiltext{1}{Based on observations obtained with the W.~M.~Keck
Observatory, which is jointly operated by the California Institute of
Technology and the University of California.}
\altaffiltext{2}{Space Telescope Science Institute, 3700 San Martin Dr.,
Baltimore, MD 21218}
\altaffiltext{3}{Astronomy Dept., Columbia University}
\altaffiltext{4}{Physics Dept., University of California--Davis}
\altaffiltext{5}{IGPP/Lawrence Livermore National Laboratory}
\altaffiltext{6}{National Optical Astronomy Observatory}
\altaffiltext{7}{Laboratory for Astronomy and Solar Physics, NASA Goddard
Space Flight Center}

\begin{abstract}
We have constructed a sample of quasar candidates by comparing the {\it
FIRST} radio survey with the 16 deg$^2$ Deeprange $I$-band survey
carried out by Postman et al.\ (1998, 2002). Spectroscopic followup of
this magnitude-limited sample ($I<20.5$, $F_\nu(20{\rm cm})>1$~mJy)
has revealed 35 quasars, all but two of which are reported here for the
first time.  This sample contains some unusual broad absorption
line quasars, including the first radio-loud FR~II BAL previously
reported by Gregg et al.\ (2000).

Comparison of this sample with the {\it FIRST} Bright Quasar survey
samples selected in a somewhat bluer band and with brighter magnitude
limits reveals that the $I$-band-selected sample is redder by 0.25--0.5
magnitudes in $B-R$, and that the color difference is not explained by
the higher mean redshift of this sample but must be intrinsic.  Our small sample
contains five quasars with unusually red colors, including three that
appear very heavily reddened.
Our data are fitted well with normal blue
quasar spectra attenuated by more than 2.5 magnitudes of extinction in
the $I$-band.

These red quasars are only seen at low
redshifts ($z<1.3$).
Even with a magnitude limit $I<20.5$, our survey is deep enough to
detect only the most luminous of these red quasars at $z \lesssim 1$; similar
objects at higher redshifts would fall below our in $I$-band limit.
Indeed, the five most luminous objects (using dereddened magnitudes)
with $z<1.3$ are {\it all} red.  Our data strongly support the hypothesis
that radio quasars are dominated by a previously undetected population
of red, heavily obscured objects.
Unless highly reddened
quasars are preferentially also highly luminous, there must be an even larger,
as yet undiscovered, population of red quasars at lower luminosity.
We are likely to be finding only the
most luminous tip of the red quasar iceberg.

A comparison of the positions of the objects in our sample with the
catalog of Deeprange cluster candidates reveals that five of our
six $z<1$ quasars are associated with cluster candidates of similar
estimated redshifts.  This association is very unlikely to be the
result of chance.  It has some surprising implications, including the
possibility that up to half of the Deeprange clusters at $z\sim1$  have
associated quasars.

\end{abstract}


\section{Introduction}

While the term quasar was introduced to describe a new population of
`QUAsi-StellAr Radio' sources, the distinction between this class and
the radio-silent `Quasi-Stellar Objects' (QSOs) with similarly extreme
optical luminosities was quickly lost. Over the first 35 years of
quasar/QSO surveys, just over 10,000 such objects were cataloged. The
majority of these were optically selected, either using objective prism
surveys to identify prominent emission lines or color selection based
on the typically blue QSO spectra; most of the remainder of the
cataloged cohort came from the optical identification of radio sources
discovered in surveys with relatively bright flux density limits. A
small additional set of objects came from identification of
serendipitous X-ray sources in the fields observed by the Einstein
(e.g., Stocke et al.\ 1983) and ROSAT (e.g., Bade et al.\ 1995; Mason
et al.\ 2000) soft X-ray telescopes.  While consensus has been reached
on the quasar energy source (accretion), their underlying agents
(supermassive black holes), and their evolution (a marked peak in the
redshift range $1<z<2$), questions remain about the completeness of the
current samples. Does a radio-loud/radio-quiet dichotomy exist, or have
existing surveys selected against radio-intermediate objects (cf.
Helfand et al.\ 1999a, White et al.\ 2000, and
Cirasuolo, Magliocchetti, Celotti \& Danese 2003 
with Ivezi\'c et al.\ 2002)?
Have we missed a large, perhaps dominant segment of the population as a
consequence of dust obscuration, or are current samples reasonably
complete (Webster et al.\ 1995; Srianand \&
Kembhavi 1997; Benn et al.\ 1998; Kim \& Elvis 1999;
Corbin et al.\ 2000)?

By the fortieth anniversary of the identification of 3C273 (Schmidt
1963), the number of known quasars will be rapidly approaching 100,000.
This explosion in the discovery rate is driven by two massive optical
surveys (2DF -- Boyle et al.\ 2000 and SDSS -- Schneider et al.\ 2002).
The additional colors and more sophisticated selection of candidates in
these surveys may help mitigate the problem of selection biases that
plagued earlier optically selected samples (Meyer et al.\ 2001). In
addition, however, new surveys in other wavelength regimes are
extending the range of quasar parameter space explored.  Our {\it
FIRST} Bright Quasar Survey (Gregg et al.\ 1996; White et
al.\ 2000; Becker et al.\ 2001; hereafter FBQS1, FBQS2, and FBQS3, respectively)
has added over 1000 quasars based on a
radio survey with a flux density limit a factor of fifty lower than
those used in the past. The 2MASS near-infrared sky survey is beginning
to be used as the basis for quasar surveys sensitive to highly reddened
objects (Gregg et al.\ 2002; Cutri et al.\ 2001; Lacy et al.\ 2002),
and the Chandra and XMM X-ray observatories are providing the first
look at the hard X-ray sky with sufficient angular resolution to
identify thoroughly buried quasar candidates (e.g., Stern et al.\ 2002;
Norman et al.\ 2002). While it is improbable that any of these new
surveys will alter fundamentally our quasar paradigm, they are likely
to have a significant impact on our understanding of the evolutionary
history of the population, our detailed models for the quasar engine,
and our use of these objects in cosmology and cosmogony.

We present here a modest sample of new quasars discovered in a program
to identify 20~cm radio sources in the 16 deg$^2$ Deeprange $I$-band
survey carried out by Postman et al.\ (1998, 2002).  The primary
motivation for examining the Deeprange fields for quasars arises in
the quest for previously overlooked quasar populations.  Most quasar
surveys to date have relied on the relatively blue quasar colors to
identify them in optical images of the sky. This reliance on color was
a pragmatic decision designed to increase the yield of quasar surveys;
i.e., the color cut drastically reduces contamination of the sample by
stars. There are other filters that can be used to achieve the same
goal, however. In the FBQS, for example, the rationale is that, while 10-20\%
of quasars are radio emitters at the 1~mJy level, almost no stars are
this radio bright. Even so, in carrying out the FBQS, we have found it
expedient to impose
a weak color cut ($O-E<2$) to eliminate galaxies from the sample. This
was necessary because the reliable division of objects into stellar and
nonstellar samples is problematic when using photographic plate-based
catalogs. The Deeprange survey affords an opportunity to improve on
this vital classification owing to the much higher (CCD) image quality.
With better classification, we can afford to eliminate all color cuts
and hence create a more unbiased sample. Furthermore, since the
Deeprange images are taken in $I$ band, the resulting quasar sample is
less biased against red objects; even without a color cut, red objects
are less likely to show up in a magnitude-limited sample drawn from
blue plate material.

In section 2 we describe the radio and optical databases employed and
discuss the {\it FIRST} Deeprange Quasar (FDQ) sample construction. Section 3
presents the results of a spectroscopic followup program in which
thirty-five new quasars and a number of other source types were
identified. A comparison of this $I$-band selected sample with the FBQS,
its implications for that sample's completeness, and striking evidence for a
large, dust-reddened, and previously overlooked population of quasars is
presented in Section 4. A discussion of the correlation between the quasars and
Deeprange cluster candidates is found in (\S5); a summary of our findings (\S6)
concludes our report.

\section{An I-band-selected Radio Sample of Quasar Candidates}

\subsection{The Deeprange $I$-Band Survey}

As part of a program to measure the galaxy correlation function at high
redshift and to detect distant galaxy clusters, Postman et al.\ (1998,
2002) conducted a deep $I$-band survey of a 16 deg$^2$ region at high
Galactic latitude, selected to have low Galactic extinction and IRAS
cirrus emission, a low \ion{H}{1} column density, and an absence of bright
stars and nearby rich galaxy clusters. The $4^{\circ} \times 4^{\circ}$
region centered at RA(J2000) $= 10^{\rm h} 13^{\rm m} 28^{\rm s}$
DEC(J2000) $= +51^{\circ} 36^{\prime} 44^{\prime\prime}$ was observed
in 1994--1996 with the prime-focus CCD camera of the KPNO\footnote{The
National Optical Astronomy Observatories are operated by AURA, Inc.,
under cooperative agreement with the National Science Foundation.} 4-m
telescope. Each of the 256 exposures was 900~s in duration and reached a
$5\sigma$ limiting magnitude of $I_{AB}=24$; the zero point is constant
over the survey area to $\lesssim 0.04$ mag.

The data were reduced in the standard manner, and a catalog of over
700,000 galaxies (and 200,000 stellar objects) was constructed using a
modified version of the FOCAS package (see Postman et al.\ 1998, 2002
for details). The final catalog provides a homogenous, calibrated set
of $I$-band source counts over the range $13\le I \le23.5$, which we have
used for comparison with our radio catalog.

\subsection{The FIRST Survey}

We have been constructing Faint Images of the Radio Sky at Twenty-cm
since the spring of 1993 using the Very Large Array (VLA)\footnote{The
National Radio Astronomy Observatory is a facility of the National
Science Foundation operated under cooperative agreement by Associated
Universities, Inc.} in its B configuration (Becker, White \& Helfand
1995).  Over 9000 deg$^2$ of the North Galactic Cap have now been
imaged to a 20~cm flux density limit of 1.0 mJy; the $\sim810,000$ radio
sources detected have positional accuracies of $\leq1^{\prime\prime}$
(White et al.\ 1997).  In particular, {\it FIRST} covers the entire
Deeprange survey area; a total of 1206 {\it FIRST} catalog sources fall
within the area of Deeprange coverage (excluding objects within
areas deleted because of bright stars, etc.). A detailed analysis of
the radio-optical comparison is in preparation; here we concentrate on
the radio sources with stellar counterparts in order to construct a
candidate quasar sample.

\begin{figure*}
\epsscale{0.95}
\plottwo{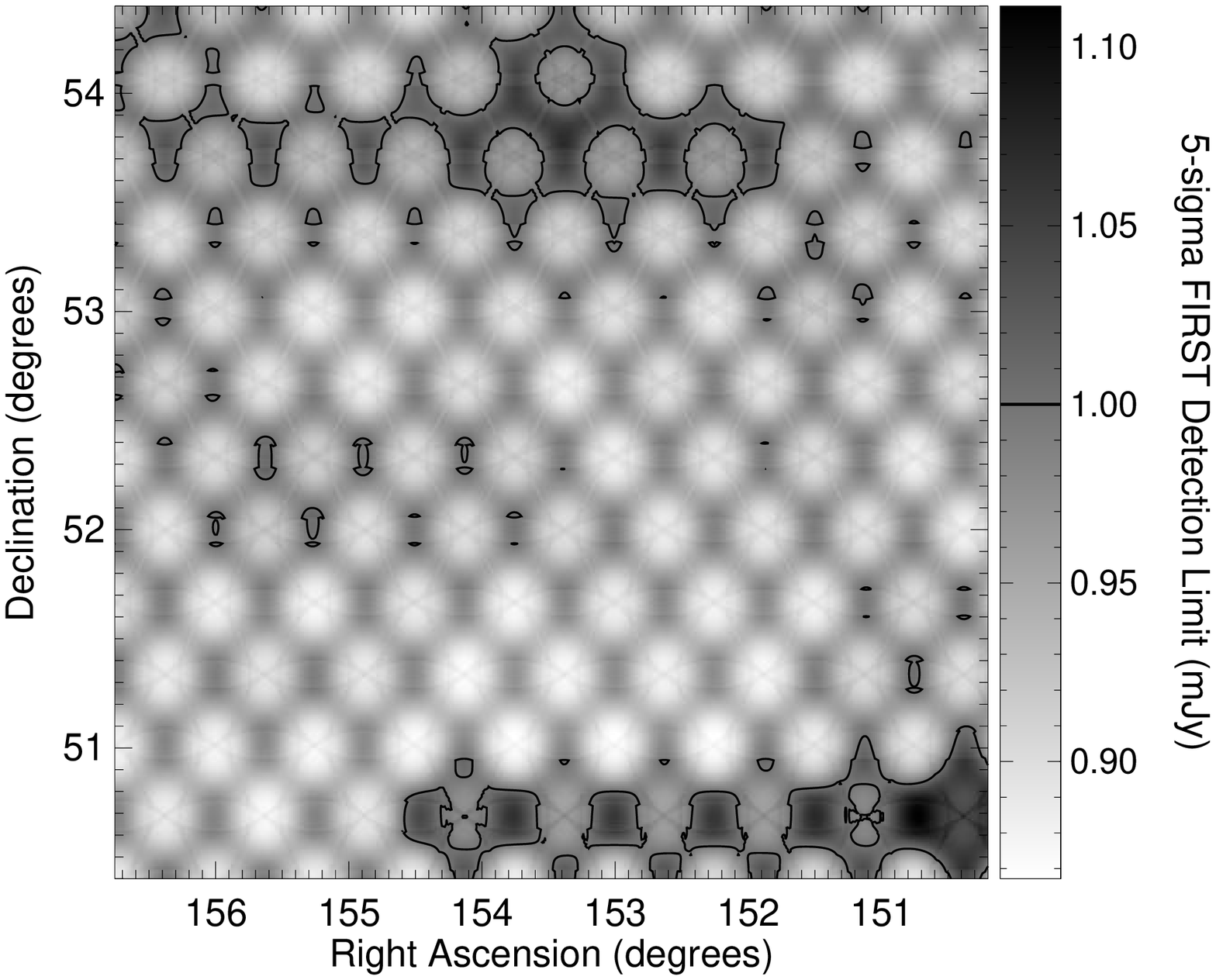}{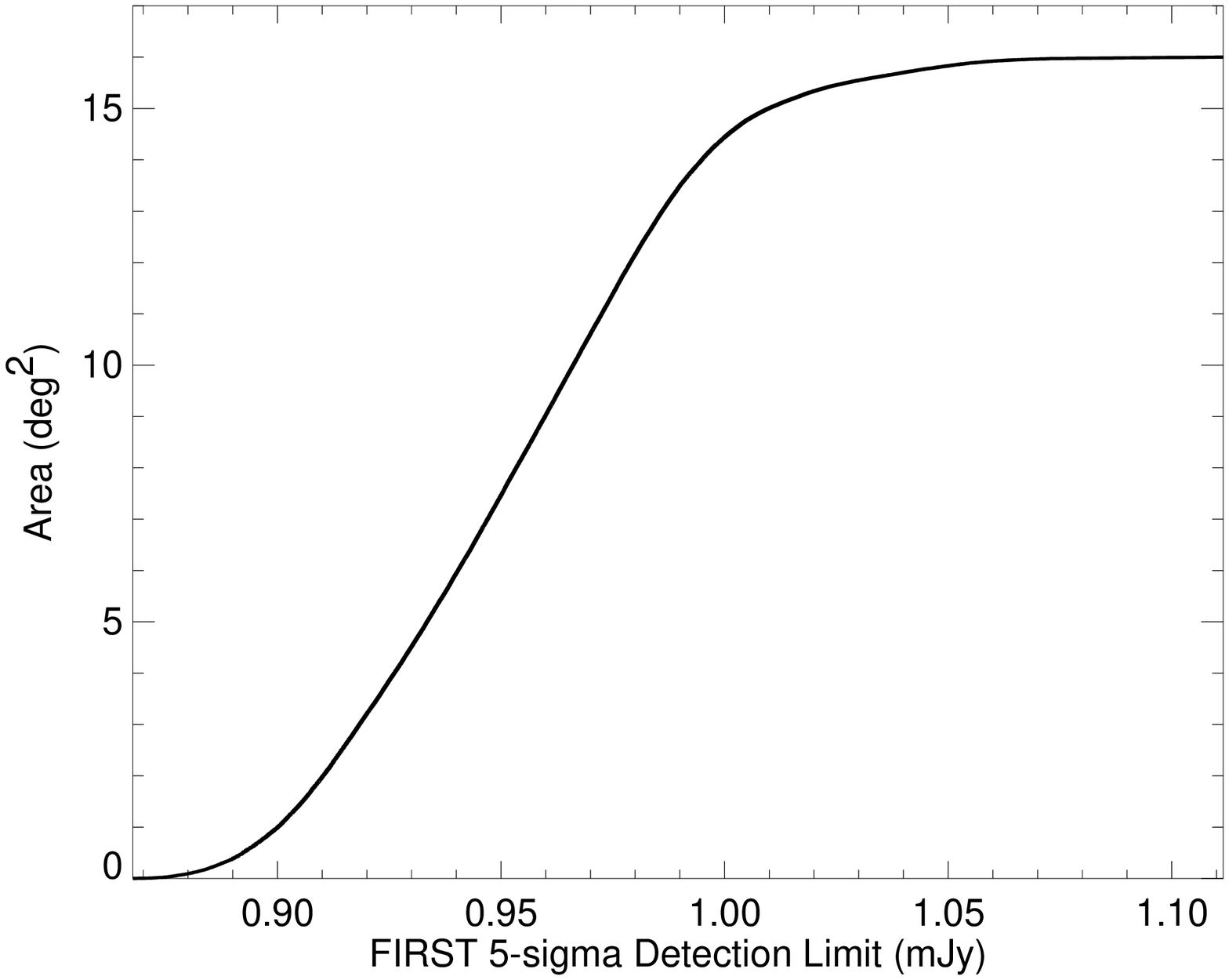}
\epsscale{1}
\caption{(a) Sensitivity of the {\it FIRST} survey as a function of position over
the Deeprange area.  The sensitivity varies slightly due to the
hexagonal pointing pattern of the survey grid and due to the
presence of bright sources that raise the noise level in the maps (see
Becker, White \& Helfand 1995 for more details.)  A contour is drawn at
1~mJy, the flux density limit for the {\it FIRST} catalog.  (b) Area covered
by {\it FIRST} survey as a function of the $5\sigma$ 1.4~GHz detection
limit.  More than 90\% of the Deeprange survey area is complete to
1~mJy, and 98.9\% is complete to 1.05~mJy.
}
\label{fig-coverage}
\end{figure*}

The {\it FIRST} survey sensitivity is quite uniform over the Deeprange survey
area.  Figure~\ref{fig-coverage} shows the {\it FIRST} survey sensitivity as
a function of position and the cumulative area covered as a function of
the {\it FIRST} $5\sigma$ detection limit.

\subsection{A Sample of Stellar Counterparts}

We began our comparison of the radio and $I$-band data by deriving
astrometric offsets between the two databases. Since the {\it FIRST}
positions are tied to the ICRF (International Celestial Reference
Frame; Ma et al.\ 1998) with an overall systematic bias of $<30$ mas
(White et al.\ 1997), we adopt them as defining the reference frame for
the match. A catalog of 978 radio/optical sources was constructed by
matching the Deeprange catalog to both the {\it FIRST} catalog and a
supplemental catalog based on VLA A-configuration observations of the
Deeprange area.  We corrected zero-point shifts in the astrometry
for each CCD and derived a distortion correction for the intra-CCD
positions in a manner analogous to our derivation of intraplate
corrections for the APM survey (McMahon et al.\ 2002); these
corrections were less than $1^{\prime\prime}$ peak-to-peak.  Both
stellar and non-stellar optical objects were included in the
calibration; non-stellar objects have poorer positions but are much
more frequent radio source counterparts, so they improve the overall
calibration.  The final registration of the two catalogs is accurate to
$\sim 0.05^{\prime\prime}$ or better, and the rms scatter for
individual sources is $1^{\prime\prime}$.

\begin{figure*}
\plotone{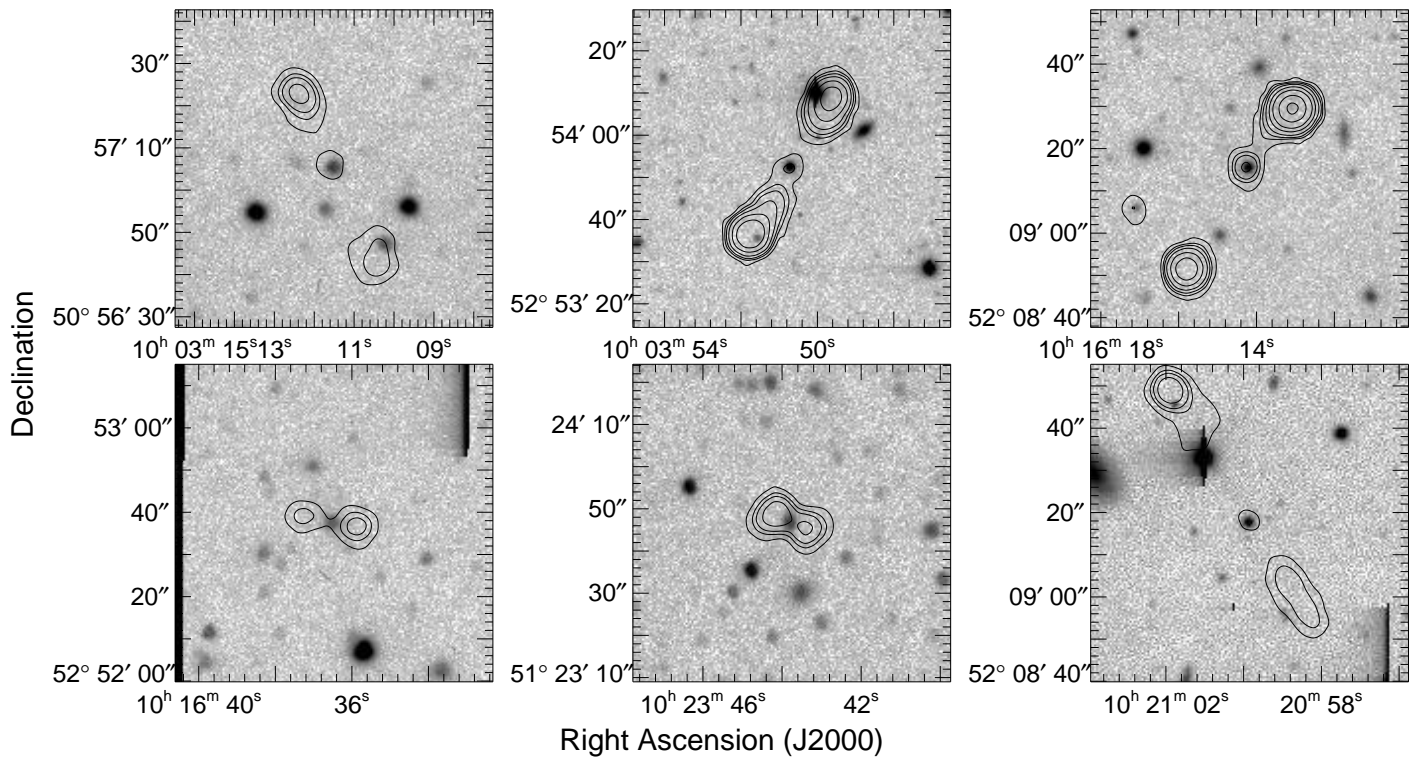}
\caption{{\it FIRST} radio contours overlaid on Deeprange $I$-band images for six
multiple component radio sources, including four quasars from Table~1
and two galaxies (bottom left and center panels) from Table~2.  The
contour levels are 1, 2, 3, 5, 10, 20, 50 and 100 mJy.  Three of these
objects (including the quasar at top center) would have been missed
using a search algorithm that simply compares positions of the {\it FIRST}
and Deeprange catalog sources.
}
\label{fig-frii}
\end{figure*}

For the vast majority of the radio sources, we search for $I$-band
counterparts by simply matching the positions. However, for
multiple-component and extended radio sources, blind matching can lead
to missed counterparts. Thus, maps were prepared for all objects by
overlaying the radio contours on the $I$-band images; the maps were then
examined by eye to select additional counterparts. Examples of these
maps for the six classical Fanaroff-Riley Type II radio doubles (FRII
-- Fanaroff \& Riley 1974) are shown in Figure~\ref{fig-frii}. Four of
these objects are quasars and two are galaxies.
In three of these cases (including one
quasar), a blind catalog match would not have found the counterpart to
lie within $1.5^{\prime\prime}$ of an $I$-band source owing to the
shifting of the centroid of the fitted Gaussian components by
emission from along the radio jet, even though an obvious (and now
confirmed) counterpart is present. Such cases are a source of modest
incompleteness in any radio identification program that does not
examine radio source morphology in detail (see, for example, Becker et
al.\ 2001 for a test of the incompleteness of the FBQS arising from
extended radio morphologies).

Of the 617 {\it FIRST} radio sources with $I$-band counterparts coincident to within
$1.5^{\prime\prime}$, 122 are classified as stellar. We calculate the
expected chance coincidence rate using the number of stellar sources in
an annulus between $5^{\prime\prime}$ and $12^{\prime\prime}$ around
each radio source in the catalog. Among the 49 stellar counterparts
brighter than $I=20.5$ we expect between 1 and 2 false matches. Another
source of misidentification with stellar counterparts arises from
galaxies misclassified as stars in the optical database; Postman et
al.\ (1998) give an estimate of the magnitude-dependent rate at which
this occurs. For $I<20.0$, we should expect 3 out of 41 such
misclassifications, whereas for $20<I<21$ we estimate that nearly half
of the putative stars are misclassified galaxies, and at fainter magnitudes the
situation deteriorates rapidly. From these considerations, coupled with
limitations imposed by the amount of spectroscopic confirmation time
available, we have adopted a magnitude-limited threshold for $I$-band
stellar counterparts of $I\le20.5$, leaving us 49 candidates of which
we expect approximately $15\%$ to be either false or misclassified
matches.

\section{Spectroscopic Identification of Quasars in the Deeprange Field}

Observations of the radio source counterparts were conducted over
several observing sessions at the Keck II 10-m telescope using both the
LRIS and ESI spectrographs. In general, the observing conditions were
not photometric. All of the observations were taken at the parallactic
angle to minimize slit losses.  The LRIS data were reduced and spectra
extracted using standard IRAF procedures.  ESI spectra were produced
using a combination of IRAF tasks and customized software developed by
the authors. More details on the algorithms used in the ESI software
are given in White et al.\ (2003).

We have obtained spectroscopic classifications for all of the 49 stellar
radio source counterparts with $I<20.5$; one additional source slightly
below this threshold was also observed\footnote{In addition, six more
objects with positional offsets greater than $1.5^{\prime\prime}$ were
observed; all were either stars (i.e., chance coincidences) or
galaxies.}. We find 35 quasars (see Table 1), one BL Lac object, two
narrow-line AGN, six galaxies with \ion{H}{2}-like spectra, five
absorption-line galaxies, and one star. The definitions of the various
classes are taken from FBQS2 and FBQS3: we define anything with broad
lines as a quasar, objects with narrow emission lines as
\ion{H}{2}~galaxies or AGN depending on their relative line strengths,
and the single object with a featureless continuum as a BL~Lac.  One or
more of the absorption-line galaxies could well contain nonthermal
components (e.g., as do BL Lacs), but our discovery spectra are
insufficient for quantitative classification.

\begin{figure*}
\plotone{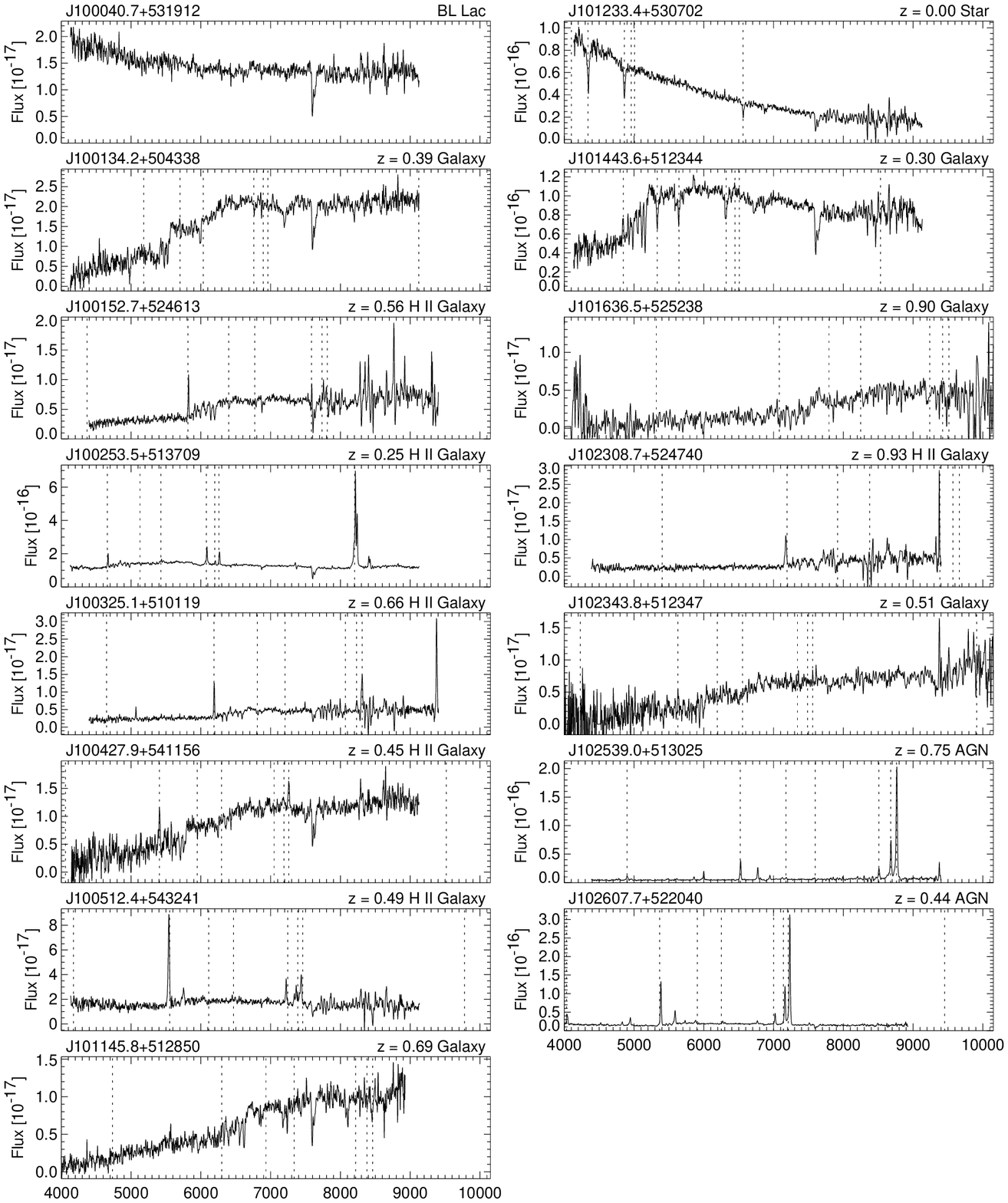}
\caption{
Spectra of {\it FIRST}/Deeprange candidates identified as non-quasars, sorted by
Right Ascension.  The labelling is the same as Fig.~3.
}
\label{fig-nonqspectra}
\end{figure*}

Since only a tiny fraction of {\it FIRST} sources are actually
identified with stars (Helfand et al.\ 1999b), the single stellar match almost
certainly represents the expected one false coincidence (although the
spectrum is unusually blue for a randomly selected star.) Likewise, the five
galaxies along with some of the six \ion{H}{2} galaxies comprise the expected
number of $I$-band misclassifications, although in this case, they almost
certainly {\it do} represent the optical counterpart to the
radio source. The BL Lac, AGN,
and the more luminous \ion{H}{2} galaxies have sufficiently bright nuclei
that their classification as stellar is to be expected. We summarize
the optical and radio properties of the non-quasar sample members in
Table 2. The spectra for the quasars and non-quasars are displayed in
Figures~\ref{fig-spectra} and~\ref{fig-nonqspectra}, respectively.

\begin{figure*}
\plotone{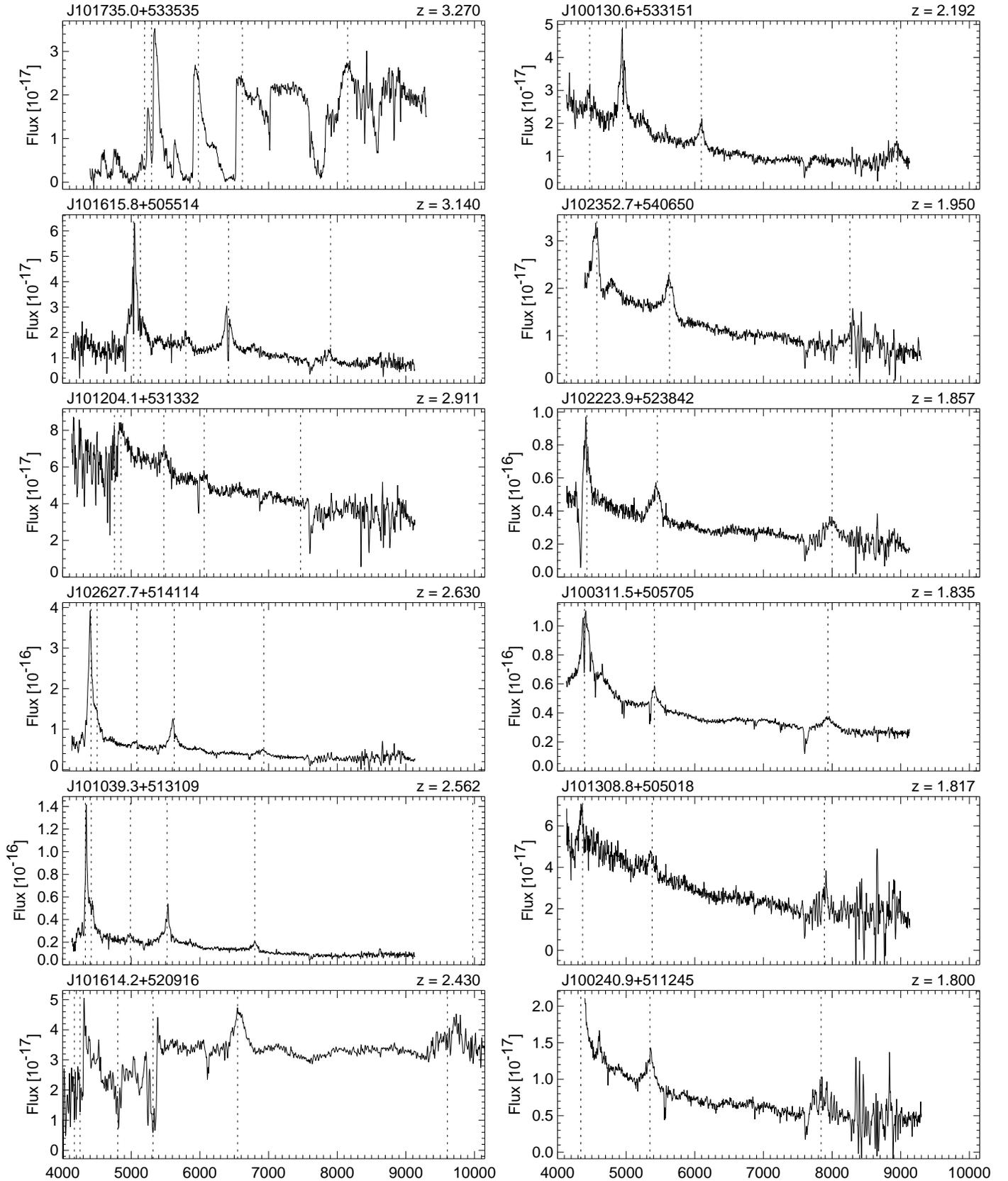}
\caption{
Spectra of {\it FIRST}/Deeprange candidates identified as quasars, sorted by
decreasing redshift.  The dotted lines show expected positions of
prominent emission lines:
Ly$\alpha$~1216,
N~V~1240,
Si~IV~1400,
C~IV~1550,
C~III]~1909,
Mg~II~2800,
[O~II]~3727,
H$\delta$~4102,
H$\gamma$~4341,
H$\beta$~4862,
[O~III]~4959,
[O~III]~5007,
H$\alpha$~6563.
Note that some of the spectra have atmospheric
A and B band absorption at $\sim6880$~\AA\ and 7620~\AA.
}
\label{fig-spectra}
\end{figure*}

\begin{figure*}
\figurenum{3b}
\plotone{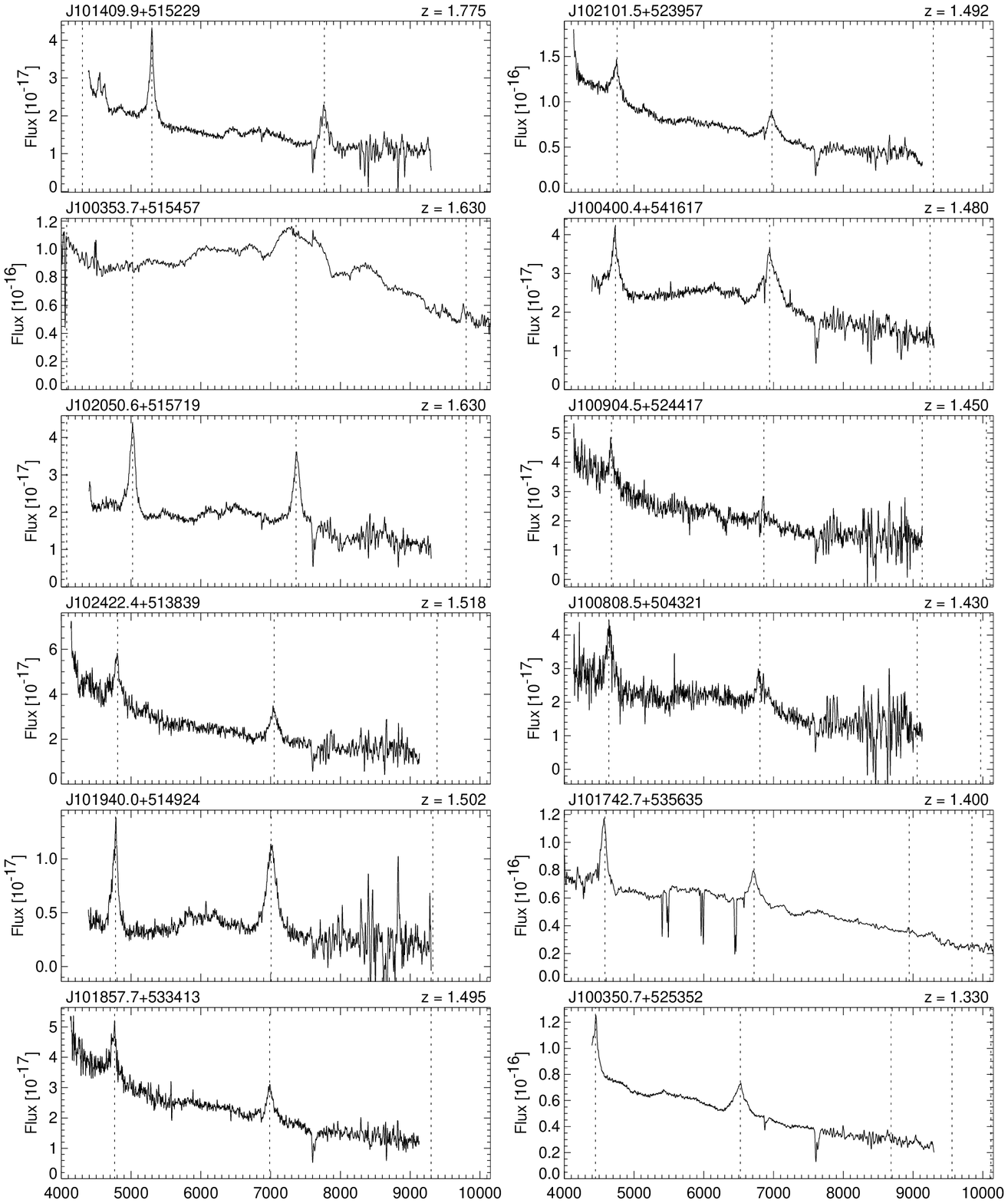}
\caption{{\it Continued.} Spectra of {\it FIRST}/Deeprange quasars.}
\end{figure*}

\begin{figure*}
\figurenum{3c}
\plotone{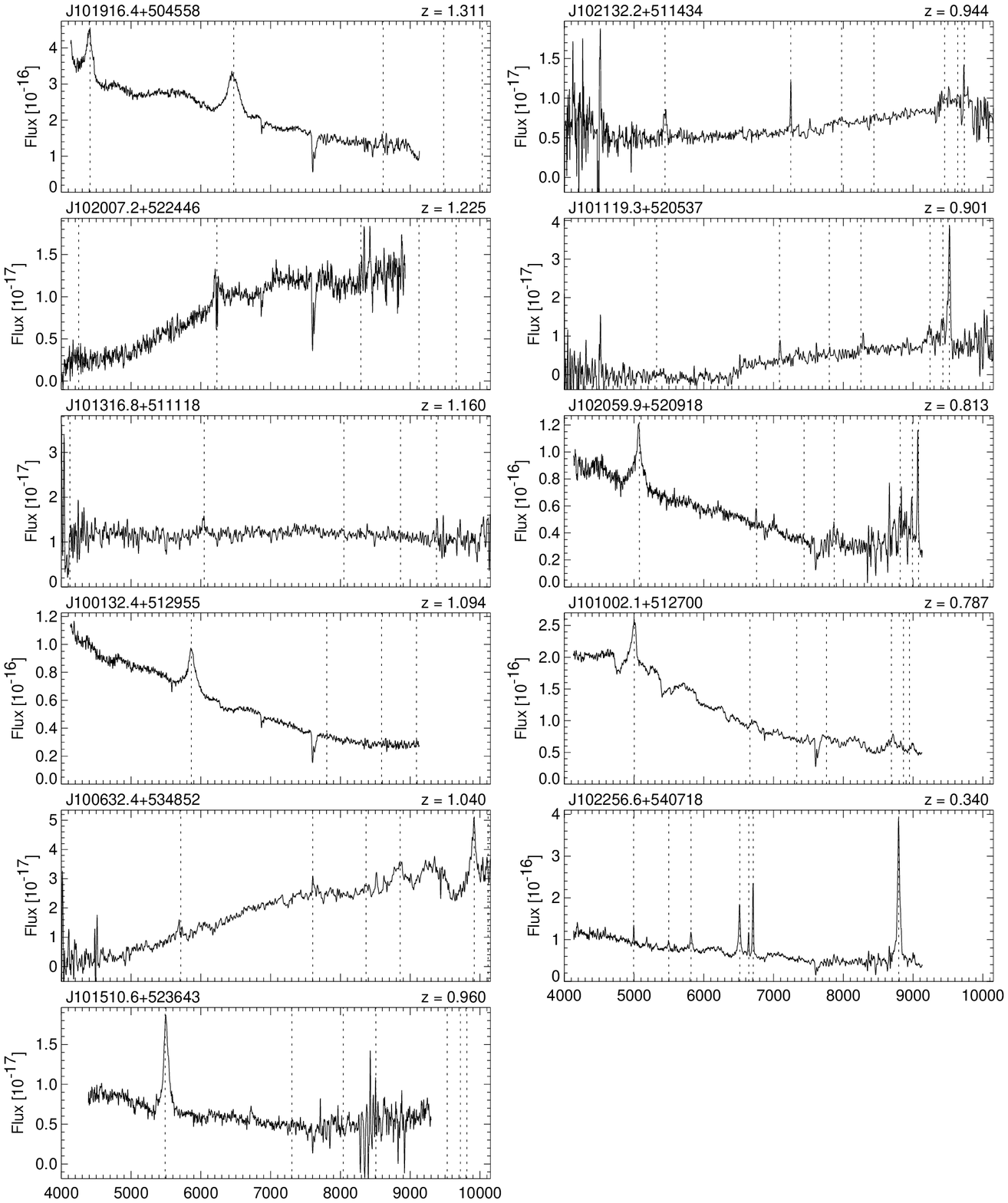}
\caption{{\it Continued.} Spectra of {\it FIRST}/Deeprange quasars.}
\end{figure*}

Table 1 lists the parameters of the Deeprange radio quasar sample. The
source coordinates from the Deeprange catalog are given in the first
two columns, followed by the offset between the optical and radio
positions. The peak 20~cm flux density of the source's core component
from the {\it FIRST} catalog and the total source size follow (col.~4
and 5). Major and minor axes derived from elliptical Gaussian fits are
quoted if one component was sufficient to describe the source
morphology; sources with major axes smaller than $2.5^{\prime\prime}$
are unresolved in the {\it FIRST} survey and their sizes are listed as
upper limits. For multiple component sources, the full source extents
were measured from the radio maps and are listed with a $``\sim``$
sign; classical FR II sources are marked with a T (for triple) or D
(for double). The next column lists the {\it FIRST} catalog integrated
flux densities for resolved single-component sources, and the sum of
the flux densities for all multiple-component sources. (For unresolved
sources, the peak and integrated flux densities are approximately
equal, although noise can produce derived integrated flux densities
somewhat below the peak values for weak sources; in these cases, the
peak values are more reliable estimators.)

Column 7 lists the 20~cm flux density reported in the NRAO VLA Sky
Survey (NVSS -- Condon et al.\ 1998). The $45^{\prime\prime}$ beam of
this survey is sensitive to low surface brightness emission that could
be resolved out in the {\it FIRST} survey. Twenty-seven of the 35 quasars
are detected in the NVSS; of the eight that lie below that survey's
threshold of 2.5 mJy, we examined the contour maps at the source
locations and found four present at flux density levels between 1 and
2~mJy (indicated as, e.g., $\sim1.4$). Comparing (as appropriate) the
peak or integrated flux densities from {\it FIRST}  with the NVSS
values shows most sources agree to well within their combined
uncertainties (systematic flux density uncertainties for the two
surveys are $\sim 5\%$ in addition to the quoted statistical errors).
Three of the point sources (FDQ J101308.8+505017, FDQ J102050.6+515719,
and FDQ J102101.4+523956) could be variable by from 50\% to a factor of
two over the one- to three-year interval separating the two sets of
observations although, given that the NVSS flux densities are higher in
the first two cases, faint diffuse emission associated with the quasar
could explain the discrepancies.

Column 8 lists the $I$-band magnitude for each quasar. These are followed
by the magnitudes in the $F$ (red) and $J$ (blue) bands derived from the
Guide Star Catalog (version 2.2.1) database\footnote{The Guide Star
Catalog-II is a joint project of the Space Telescope Science Institute
and the Osservatorio Astronomico di Torino.  Additional support is
provided by European Southern Observatory, Space Telescope European
Coordinating Facility, the International GEMINI project and the
European Space Agency Astrophysics Division.}.  If the object is not
present in this catalog, we record the APM POSS I $E$ (red) and $O$ (blue)
magnitudes from the McMahon et al.\ (2002) {\it FIRST} optical
identification catalog; three objects fall below the thresholds of both
sets of plate scans and so are excluded from the color plots discussed
below. Note that the POSS and $I$-band data are far from
contemporaneous, and substantial source variability is possible,
suggesting that the $I-$POSS colors are not highly reliable.

Columns 11 and 12 give $E(B-V)$, the reddening in the quasar rest
frame, and $A(I)$, the extinction
in the observed frame for the $I$-band, as determined from a fit to
each spectrum.
They are discussed in more detail below (\S\ref{section-red}).

The quasar redshift, plus comments on the source's spectrum and its
properties in other wavelength bands, complete the Table. Only one of
the quasars (the optically brightest member of the sample, FDQ
J101916.3+504557) has been reported prior to its {\it FIRST} detection
(Stepanian et al.\ 1999); Gregg et al.\ (2000) earlier reported the
discovery of the FR II BAL quasar FIRST J101614.2+520915. Three of these
quasars are coincident with X-ray
sources detected by ROSAT (but see below).

Most of the quasar spectra are unremarkable, although a few deserve
comment:

\noindent {\bf FDQ J100353.6+515457 ($z=1.63$)} ---
This object has an unusual spectrum with very broad \ion{Mg}{2} emission
and very strong \ion{Fe}{2} emission.

\noindent {\bf FDQ J101204.0+531331 ($z=2.911$)} ---
This quasar's emission lines have substantially lower
equivalent widths than the FBQS composite spectrum (Brotherton
et al.\ 2001).

\noindent {\bf FIRST J101614.2+520915 ($z=2.43$)} ---
As was discussed in Gregg et al.\ (2000), this object is a high
ionization broad absorption line quasar (HiBAL) with an FRII radio
morphology, conclusively demonstrating that BALs can be radio loud.

\noindent {\bf FDQ J101735.0+533535 ($z=3.27$)} ---
This is a low ionization broad absorption line quasar (LoBAL) based on
the presence of broad absorption by both \ion{C}{4} and \ion{Al}{3}.

\noindent {\bf FDQ J101742.7+535635 ($z=1.40$)} ---
This quasar has very strong absorption lines due to \ion{Mg}{2} and Fe
at a redshift of $z=1.305$, indicative of an intervening
damped Lyman alpha line system (Rao \& Turnshek 2000).

\noindent {\bf FDQ J101939.9+514924 ($z=1.502$)} ---
This quasar's emission lines are unusually strong (or its continuum
weak) compared with the FBQS composite spectrum.

\noindent {\bf FDQ J102223.9+523842 ($z=1.857$)} ---
This appears to be a HiBAL as well, albeit a marginal example of one.

\noindent {\bf FDQ J100632.3+534852 ($z=1.040$)},

\noindent {\bf FDQ J101119.2+520536 ($z=0.901$)},

\noindent {\bf FDQ J101316.8+511118 ($z=1.16$)},

\noindent {\bf FDQ J102007.2+522445 ($z=1.225$)},

\noindent {\bf FDQ J102132.2+511433 ($z=0.944$)} ---
These quasars all have red spectra.  They are discussed further
below (\S\ref{section-red}).

\section{Discussion}

Our spectroscopic campaign has resulted in the discovery of 35 quasars.
One might wonder whether it is worthwhile to create a new, small sample
of quasars even as large quasar surveys from the Sloan Digital Sky
Survey and 2dF are producing samples with thousands or tens of
thousands of objects.  We believe that this and similar focused
studies continue to be interesting because they explore new parts of
parameter space.  The SDSS/{\it FIRST} sample (Ivezi\'c et al.\ 2002)
currently includes 441 spectroscopically confirmed quasars with
$i^*<18.5$ over 1030~deg$^2$ of sky.  It is therefore 2 magnitudes brighter
than this paper's sample. As a result it is less dense on the sky by a
factor of 5, and it includes only the bright end of the quasar
luminosity distribution (as does the FBQS2 sample).  Consequently the SDSS/{\it FIRST}
redshift distribution is much more heavily dominated by low-redshift objects.

On the other hand, the 2dF QSO Redshift Survey (Croom et al.\ 2001) is
deep ($18.25 < b_J < 20.85$), and includes $\sim11000$ quasars spread over
289.6~deg$^2$.  Presumably about 10\% of these would be detected by a
1~mJy radio survey like {\it FIRST} (although only the northern equatorial
strip is actually covered by {\it FIRST}); the 2dF sample thus includes about
3.8 {\it FIRST} quasars deg$^{-2}$, a slightly higher density than the
sample reported here.  But the 2dF survey is expected to be complete
only for $0.3<z<2.2$ and detects no quasars with $z>3$.  In contrast,
in our 35 object sample we detect two $z>3$ quasars.  Since it is a
survey defined from a blue magnitude, the 2dF catalog is also biased
against the inclusion of red objects. As we show below, 2dF, and most previous
quasar surveys, have missed a large population of red objects.

\subsection{Redshift Distribution}

\begin{figure*}
\epsscale{0.65}
\plotone{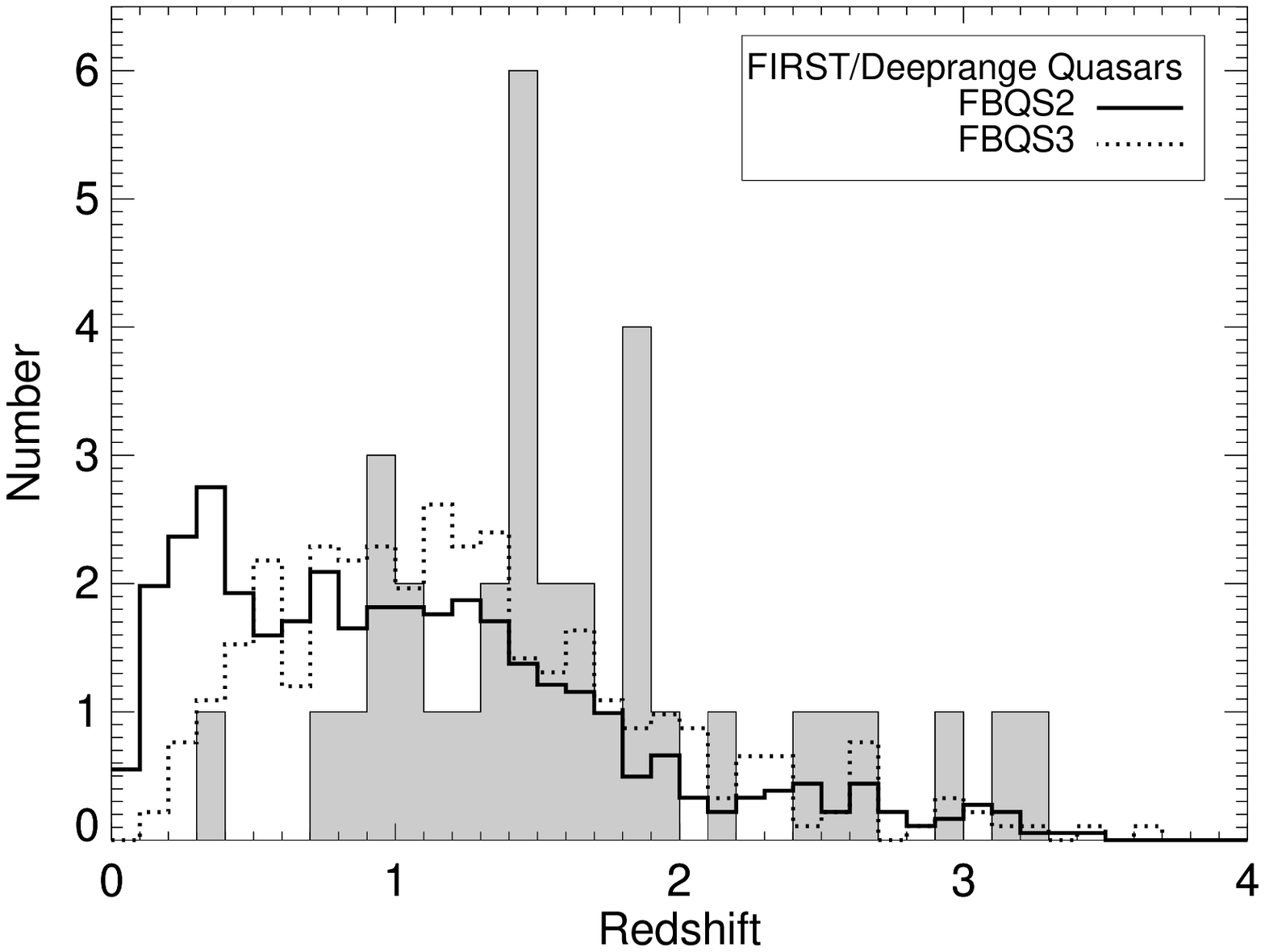}
\epsscale{1}
\caption{Redshift distribution for {\it FIRST}/Deeprange quasars (shaded)
compared with distributions for the FBQS2 (solid; White et al.\ 2000)
and FBQS3 (dotted; Becker et al.\ 2001) samples normalized to the same
total counts.  The FDQ sample contains many fewer low redshift
quasars.  The narrow peaks at $z=1.45$ and 1.85 are also striking.
}
\label{fig-zdist}
\end{figure*}

\begin{figure*}
\plotone{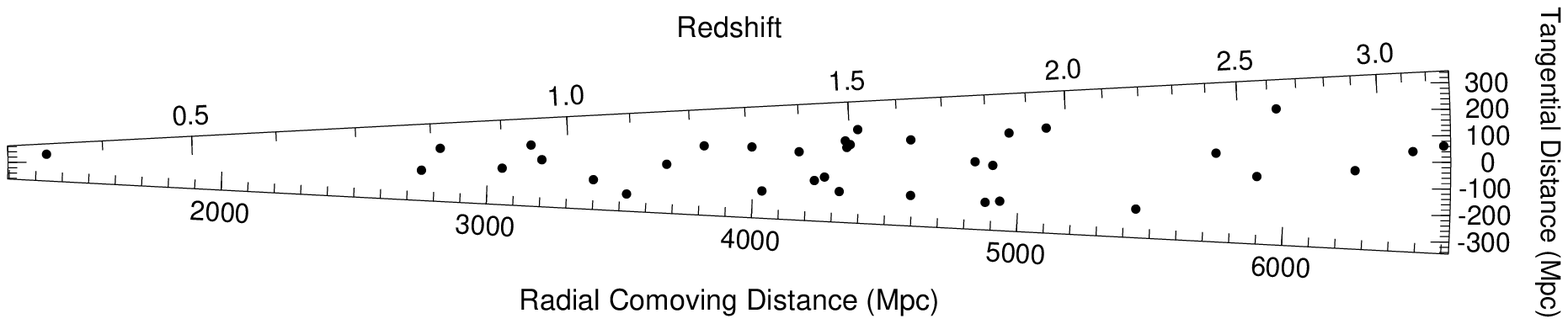}
\caption{Distribution of FDQ quasars in comoving spatial coordinates.
Distances were calculated using a cosmological model with $H_0=70$,
$\Omega_M = 0.3$, and $\Omega_\Lambda = 0.7$.  The tangential distance
is measured in the RA direction (so declination is normal to the page.)
The paucity of quasars at redshifts less than 0.8 is due both to the
limited volume covered by the survey and to the rapid increase in the
quasar number density at $z \sim 1$.
}
\label{fig-wedge}
\end{figure*}

The limiting $I$ magnitude of 20.5 for this sample is nearly 3 magnitudes
fainter than the limit in the FBQS1\&2, and 1.5 magnitudes
deeper than the limit in FBQS3. The samples do show some
distinct differences, perhaps the most pronounced being in the redshift
distribution. Continuing a trend discernible in comparisons between
FBQS2 and FBQS3, as one goes to fainter magnitudes a significant deficit
of low redshift quasars develops, to the extent that, in the Deeprange
sample, there are very few quasars below a redshift of 1 (see
Fig.~\ref{fig-zdist}). The lack of low-redshift quasars results mainly
from the small volume being sampled (Fig.~\ref{fig-wedge}) combined
with the excellent sensitivity of the survey, which is capable of
detecting even low-luminosity quasars for $1<z<2$ where there are far
more quasars than at lower redshifts.

It is possible the requirement that the $I$-band images to appear
point-like causes us to miss some lower luminosity AGN at low redshifts
because their host galaxies are also detected.  However, the number of
such objects is not expected to be very large.  From the FBQS catalog,
we expect to find $\sim1.4$ quasars with $F\le17.8$ and $z<0.7$ in the
FDQ survey area; the FBQS3 catalog predicts $\sim1.7$ quasars with
$F<18.9$.  The single $z<0.7$ quasar we detect is not much below
expectations given the slowly rising counts for low redshift AGN.

The narrow peaks at $z\sim1.5$ and 1.9 in the redshift distribution are
notable.  They are created by groups of a few objects (also visible in
Fig.~\ref{fig-wedge}) with characteristic pair separations of $\sim 60
h^{-1}$~Mpc.  Possible quasar clustering on similarly large scales has
been reported before (e.g., Graham, Clowes \& Campusano 1995; Komberg,
Kravtsov, \& Lukash 1996).  The statistical evidence for clustering in
this small sample is fairly weak using conventional measures of
clustering; we defer the complete analysis of the clustering
properties of this and other radio quasar samples to another paper.

\subsection{Broad Absorption Line Quasars}

One of the more interesting results to come out of the FBQS was the
large number of BALQSOs in the radio-selected sample. While traditional
blue-selected samples contain 10\% BALQSOs, the FBQS contains 18\%.
It is unsurprising, then, that this radio-selected Deeprange sample
also contains a relatively high fraction of BALQSOs. Out of nine quasars
at a sufficiently high redshift to allow a \ion{C}{4} BAL to be seen, three
are BALQSOs (33\%).  Although this is a small sample, it does add
weight to the conclusion that, contrary to some earlier work,
radio-intermediate quasars are more likely to be BALQSOs than radio
quiet quasars; equally important, this sample of 35 quasars includes
the first example of a radio-loud FR~II BAL quasar (Gregg et
al.\ 2000).

\subsection{Colors for Faint $I$-band Selected Quasars}
\label{section-red}

\begin{figure*}
\epsscale{0.65}
\plotone{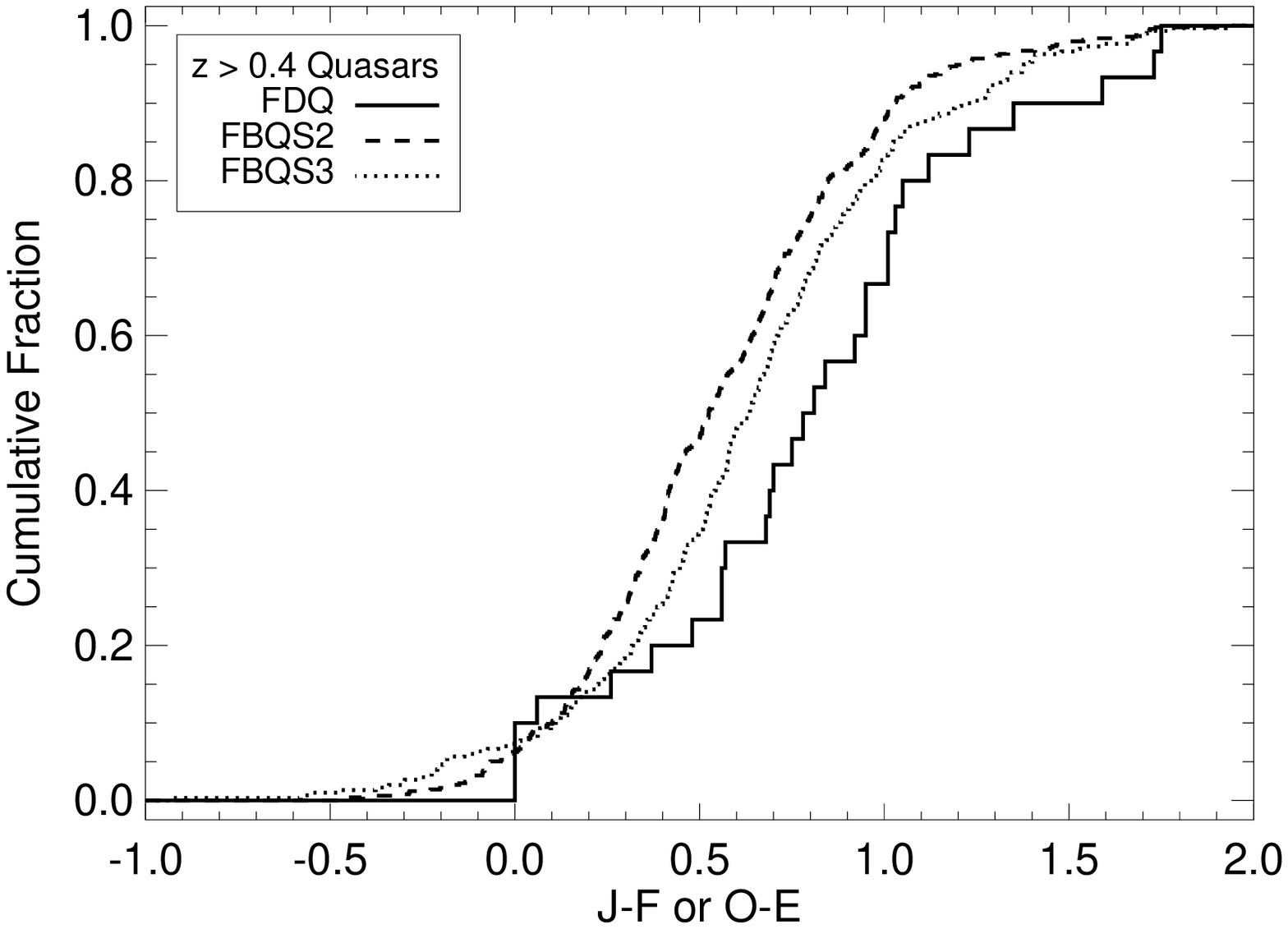}
\epsscale{1}
\caption{Cumulative color distributions for {\it FIRST}/Deeprange quasars
(solid), FBQS2 (dashed) and FBQS3 (dotted).  The FDQ sample is
noticeably redder than the other samples (which were selected using $R$
magnitudes instead of $I$ as in the FDQ sample.)
}
\label{fig-colordist}
\end{figure*}

\begin{figure*}
\epsscale{0.65}
\plotone{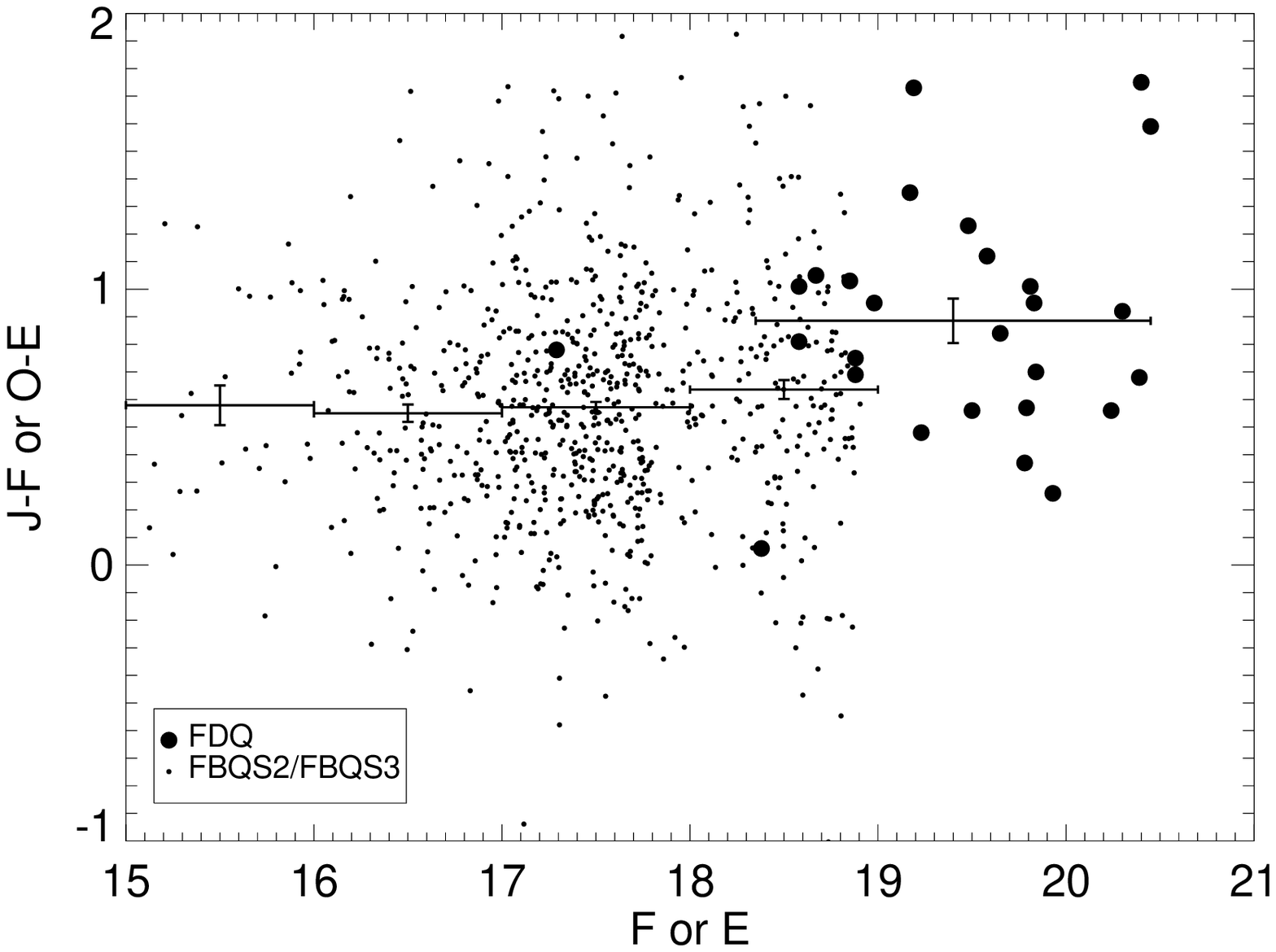}
\epsscale{1}
\caption{Colors of FDQ quasars (large symbols) and FBQS2/FBQS3 quasars
(small symbols) as a function of red magnitude.  The error bars for the
four brighter bins show the mean and error in the mean for the FBQS
quasars over the magnitude range.  The faintest bin shows the mean
for the FDQ
quasars.  There is a slight trend toward redder colors at fainter
magnitudes in the FBQS samples, but the FDQ sample is significantly redder
than the extrapolation of that trend.  Note that this diagram
omits seven FDQ sources that are undetected in one or both optical
colors; the spectra show six of them to be quite red.
}
\label{fig-colormag}
\end{figure*}

The median $O-E$ colors (roughly equivalent to $B-R$) for FBQS2 and
FBQS3 quasars with $z>0.4$ are 0.52 and 0.63, respectively, for survey
limits of $E=17.8$ and $E=19.0$. For the current survey limit of
$I=20.5$ we might expect the median to be another $\sim 0.1$ magnitude
redder, but the observed median for the 27 $z>0.4$ Deeprange quasars with both
colors shows an increase more than three times this large: $J-F=0.84$.
The comparison is displayed
graphically in Figures~\ref{fig-colordist} and \ref{fig-colormag}.  A
two-sided Kolmogorov-Smirnov test indicates that the FDQ and FBQS2
distributions are different with 99.94\% confidence, while the FDQ and
FBQS3 distributions differ with 97.8\% confidence.

\begin{figure*}
\epsscale{0.65}
\plotone{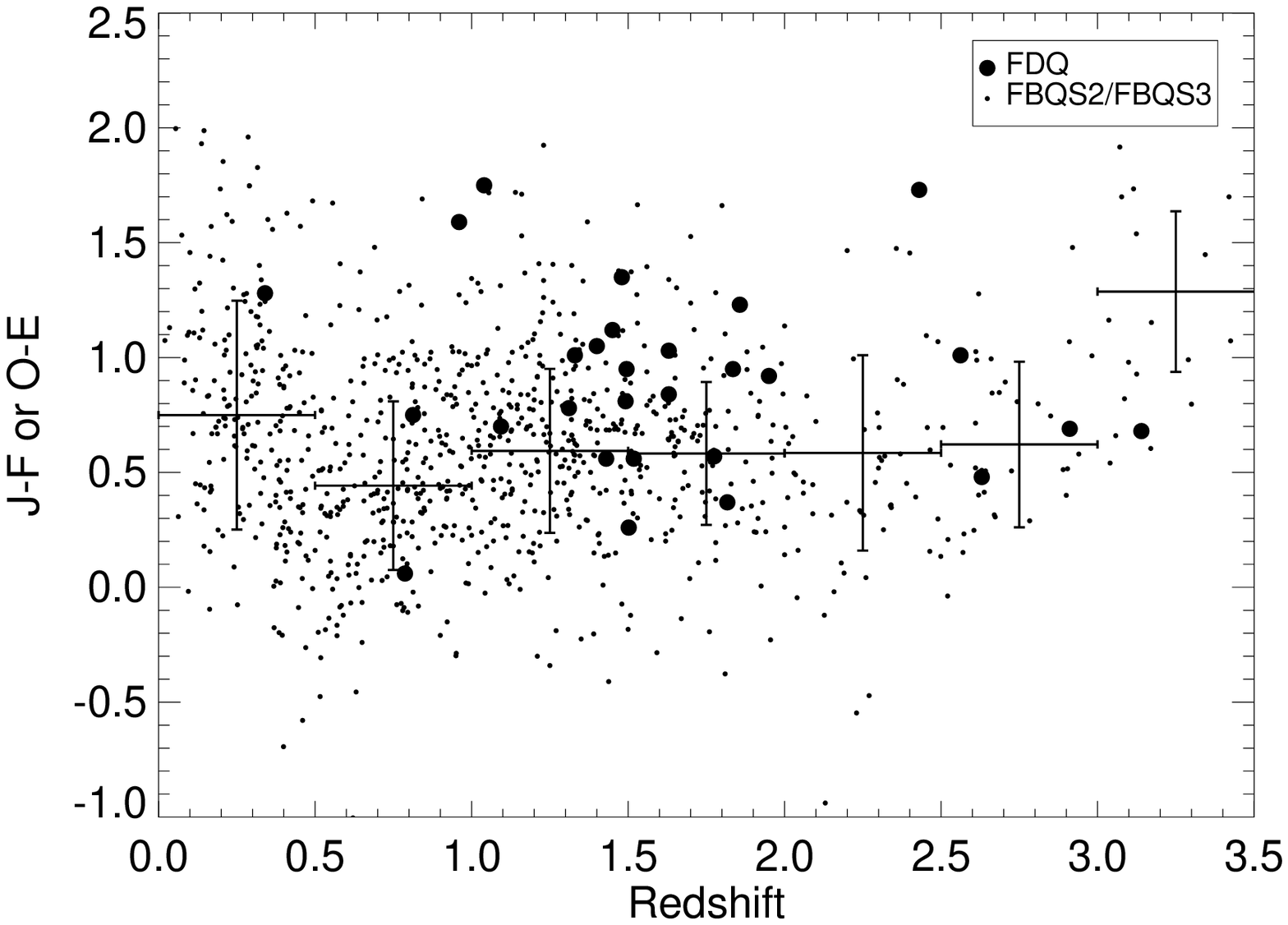}
\epsscale{1}
\caption{Colors of FDQ quasars (large symbols) and FBQS2/FBQS3 quasars
(small symbols) as a function of redshift.  The error bars show the
mean and rms for the FBQS quasars in redshift bins.  The FDQ sample is
somewhat redder than the FBQS, but the difference does not depend
strongly on redshift.
}
\label{fig-colorzdist}
\end{figure*}

The mean and median $J-F$ colors are identical for the $z<2$ FDQ
quasars and for the entire FDQ sample.  The redder colors are therefore
not a consequence of the higher mean redshift of the $I$-band sample
(Fig.~\ref{fig-colorzdist}).  The color difference is also not the
result of the various red and blue bands employed in the different
surveys.  The $J$ and $F$ bands are slightly redder than the $O$ and
$E$ bands, extending about 30~nm farther to the red in each case.
Consequently $J-F < O-E$ for the reddest objects.  Thus by comparing
the $J-F$ colors for the FDQ sample directly with $O-E$ for the FBQS
samples, we get a conservative estimate of the color difference between
the samples; the actual color difference will be larger (by
up to 0.5 mag) for the reddest objects.

It appears that the quasars are redder both because of the fainter optical
magnitude limit and as a result of being selected in a redder band. This is
consistent with the result expected if dust extinction is relevant: a
quasar of a given intrinsic luminosity with extra dust will be both
fainter and redder. Ignoring such effects can lead to a serious underestimate
of the size of the red quasar population.

In addition to the redder mean color of the sample, the removal of the
color cut (at $O-E<2.0$) has allowed us to identify three objects too
red to be included in the FBQS.\footnote{ FDQ J101316.8+511118 also is
not detected on the POSS plates, but its flat optical spectrum and weak
lines suggest that it is likely to be variable with a blazar continuum
component rather than being a red object.} One of these is a
low-ionization BAL (FDQ J101735.0+533535, $z=3.270$),
but the other two simply
have red spectra (in Fig.~\ref{fig-spectra}, see FDQ J102007.2+522446,
$z=1.225$, and FDQ J101119.3+520537, $z=0.901$.)
The $J-F$ colors for these three objects,
estimated from the spectra using synthetic photometry, are 2.3, 2.6 and
$>3$ respectively.  A fourth object with a similarly red spectrum, FDQ
J100632.4+534852 ($z=1.040$),
formally fails to exceed the color cut, but lacks
simultaneous red and blue magnitudes which would probably show that it
too would have failed to be included in the FBQS sample; its synthetic
$J-F$ is 2.5. Thus, $\sim 10\%$ of the quasars confirmed here would have
been missed in a radio-selected sample even when the color cut is more
than a magnitude redder than the sample mean.

Are these red spectra the result of extinction by dust?
To estimate the reddening for our quasars, we have fitted the spectra
using the FBQS composite spectrum (Brotherton et al.\ 2001) with a
Small Magellanic Cloud (SMC) dust reddening law from Gordon \& Clayton
(1998).  The SMC extinction rises steeply into the ultraviolet (making
it possible to produce red spectra with less visual extinction) and
lacks the 2175~\AA\ bump seen in Galactic dust.  The SMC law is similar
to dust observed in energetic extragalactic sources such as quasars
and starbursts; no
quasar has been seen to have a 2175~\AA\ absorption bump (Pitman,
Clayton, \& Gordon 2000).  We have not attempted to fit the spectra for
the two quasars that have strong broad absorption line systems, since
those objects are heavily reddened by the BALs themselves.
Table~1 gives the results determined from these fits: $E(B-V)$, the
reddening in the quasar rest frame, and $A(I)$, the $I$-band extinction
in the observed frame.

\begin{figure*}
\epsscale{0.45}
\begin{tabular}{cc}
{\plotone{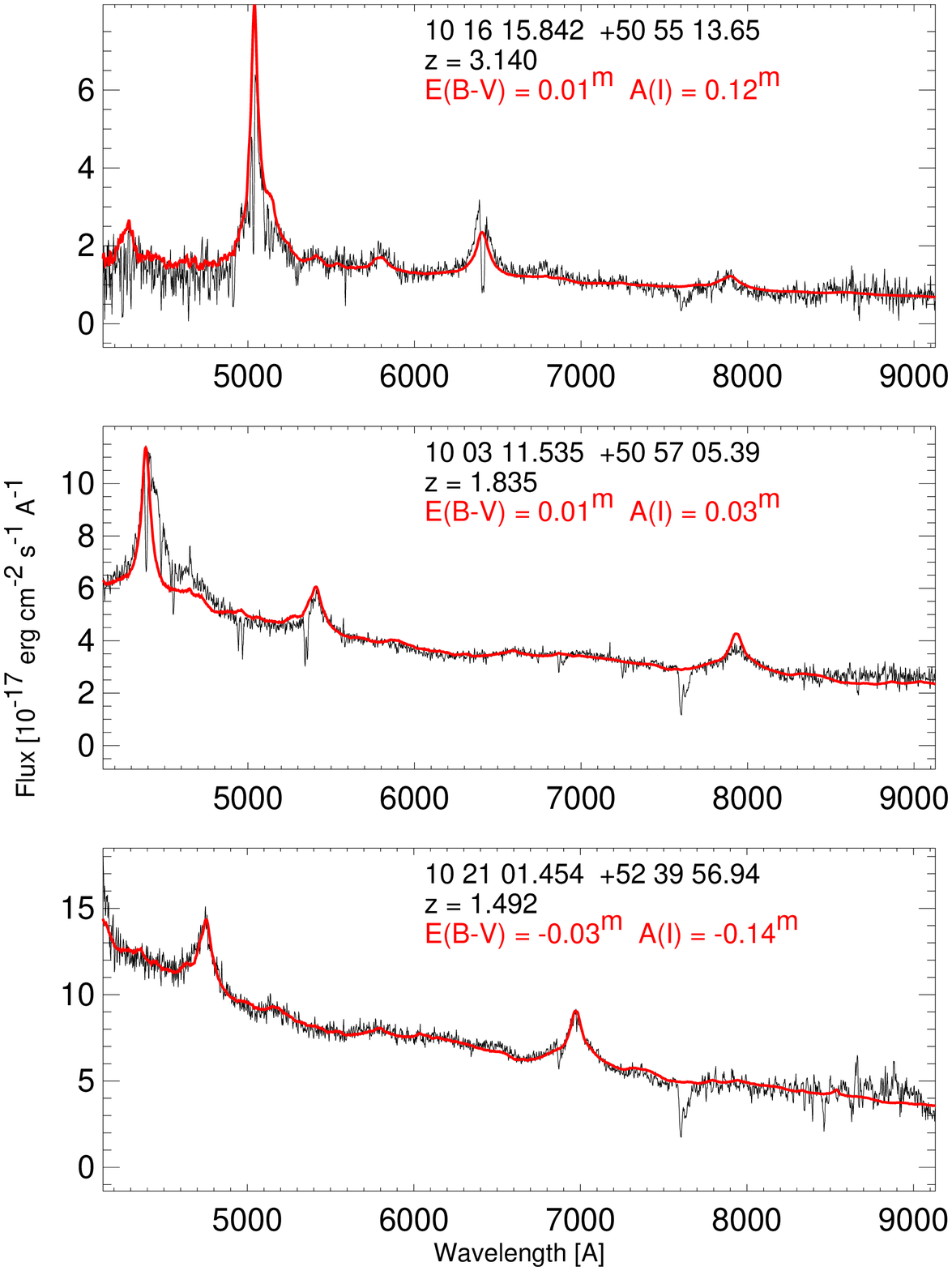}} &
{\plotone{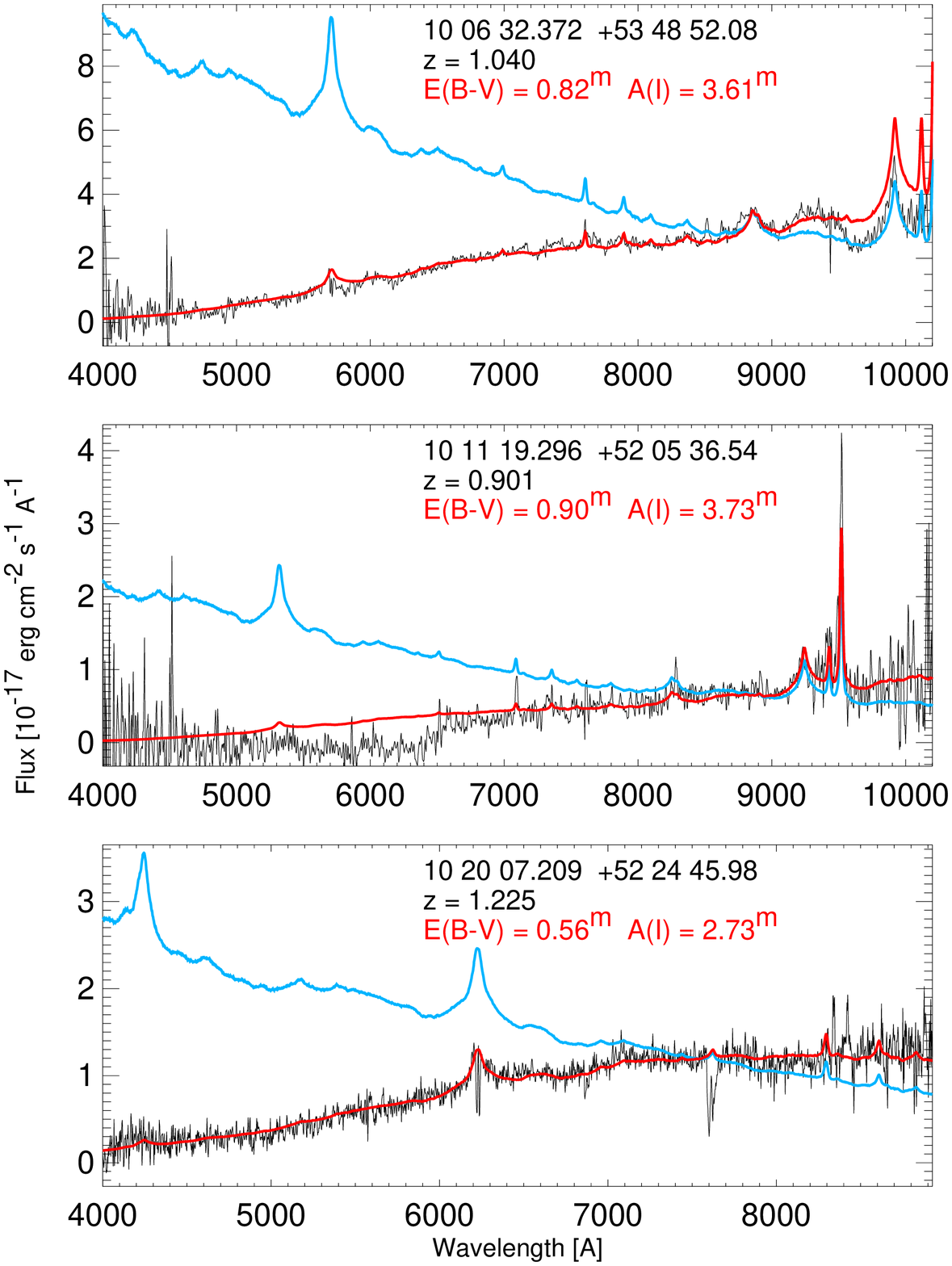}} \\
(a) & (b) \\
\end{tabular}
\epsscale{1}
\caption{
Results of fitting the FBQS composite spectrum (Brotherton et
al.\ 2001) reddened with an SMC extinction law (Gordon \& Clayton 1998)
to the FDQ spectra. (a) Fits to three selected quasars with typical
spectra having little reddening.  The $E(B-V)$ and $I$-band extinction
$A(I)$ are given.  The reddening is allowed to be negative in the fit
since the FBQS composite itself could include a small amount of
extinction. (b) Fits to the three very red quasars.  The unreddened
composite spectrum, matched to the red half of the data,
is also shown.  The fits to FDQ J100632.3+534852
(top) and J102007.2+522445 (bottom) are excellent; the fit to
J101119.2+520536 is good, but the observed spectrum has an unexplained
break at 6500~\AA.  The extinction for these quasars is large.
}
\label{fig-reddening}
\end{figure*}

\begin{figure*}
\epsscale{0.65}
\plotone{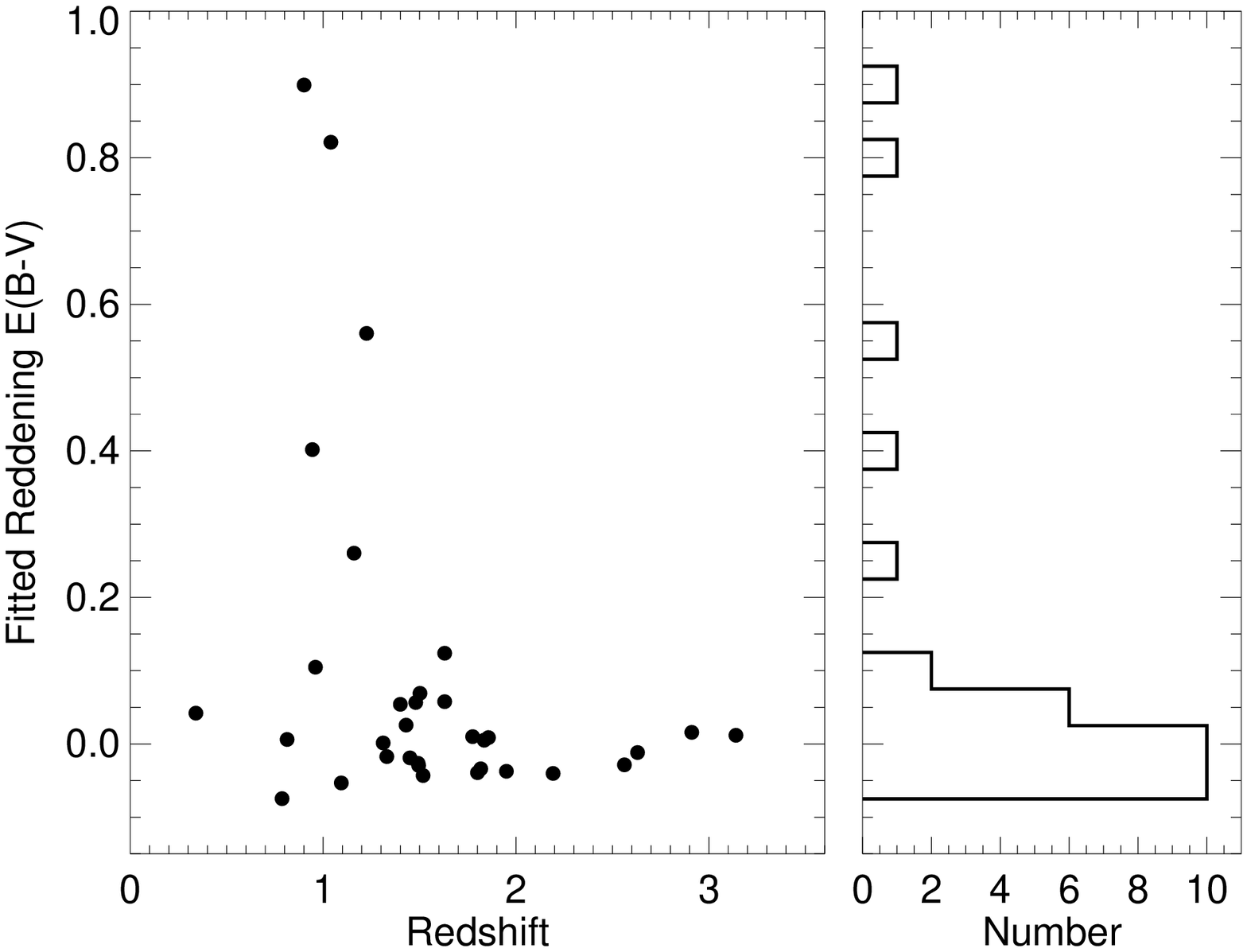}
\epsscale{1}
\caption{Distribution of fitted values for $E(B-V)$
dust reddening.  Quasars with red colors due to broad
absorption lines rather than dust have been omitted.  All other red
quasars are seen only at relatively low redshifts, $z<1.3$. The lack
of blue light in their spectra makes
more distant examples too faint to detect in the $I$-band.
}
\label{fig-ebvplot}
\end{figure*}

As can be seen from the examples in Figure~\ref{fig-reddening}(a), most
of the quasars are excellent fits to the template spectrum with little or no
reddening, but a minority show evidence for quite large extinction
values (Fig.~\ref{fig-reddening}b).  The reddened models fit the
observed spectra well; note that both the continuum and
broad emission lines are well matched by the model.

One might be tempted to assume
that because the red quasars represent only a small fraction of the
sample, they are also a small fraction of the quasar population.
However, Figure~\ref{fig-ebvplot} shows a very interesting correlation:
{\sl all\/} of the red quasars are found at relatively low redshifts.
Five of the ten quasars with $z<1.3$ have very red colors.  A two-sided
Kolmogorov-Smirnov test indicates that the redshift distributions of red
quasars (with $E(B-V) > 0.5$) and non-red quasars differ at the 96.5\%
confidence level.

\begin{figure*}
\epsscale{0.70}
\plotone{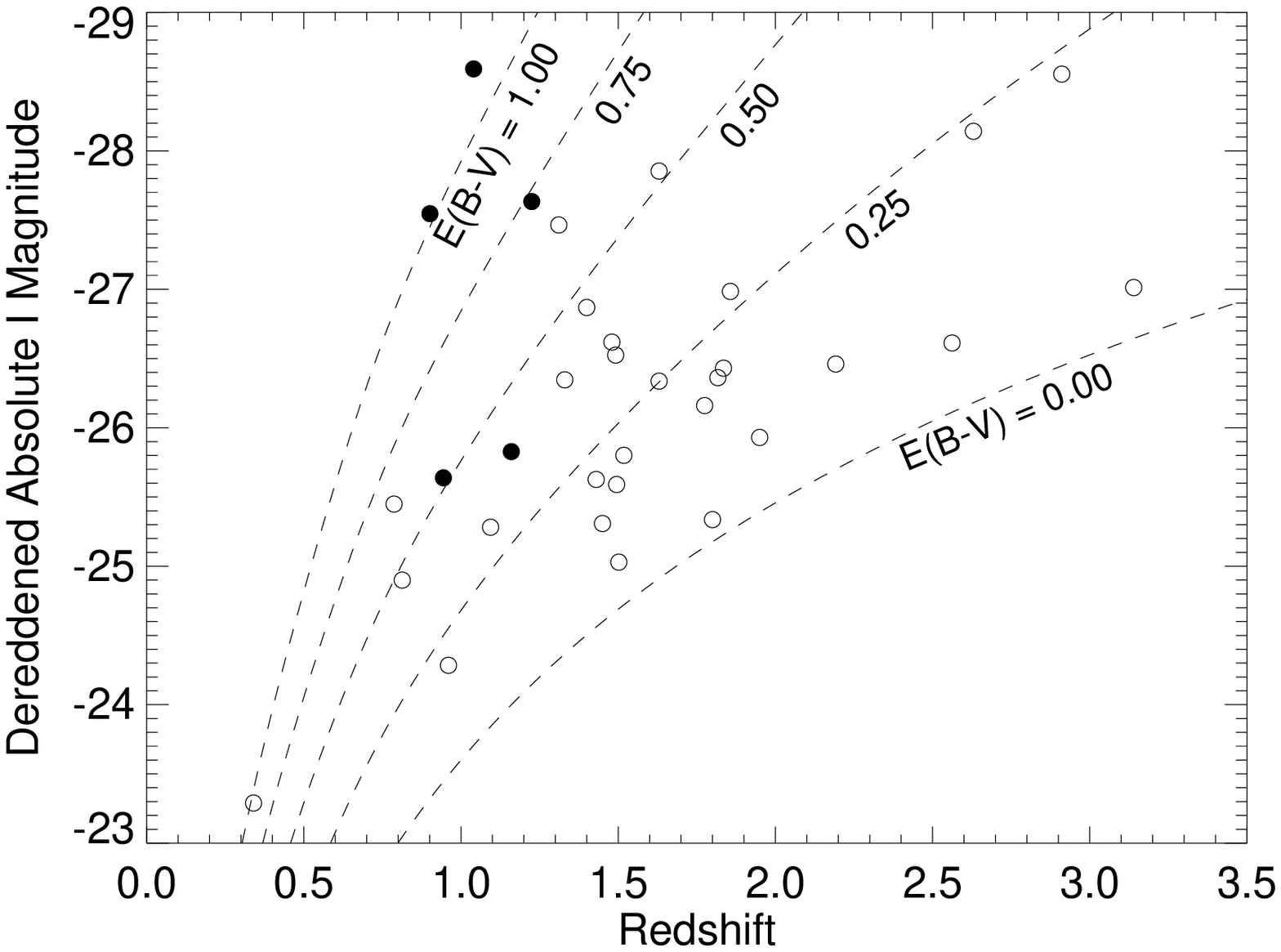}
\epsscale{1}
\caption{
Dereddened $I$-band absolute magnitudes as a function of redshift.  The
five red quasars with observed $E(B-V)>0.2$ are indicated by filled
symbols.  The dashed lines show our $I<20.5$ detection limit as a
function of the amount of reddening, $E(B-V)$.  Even the most luminous
quasars are undetectable in the $I$-band at redshifts $z>1.5$ in the
presence of $E(B-V)=0.75$ magnitudes of reddening.  We are likely to be
detecting only the most luminous tip of the red quasar iceberg.
}
\label{fig-absmag}
\end{figure*}

Red quasars are not seen at higher redshifts because their red colors
make them very difficult to detect there.  As the redshift increases
and bluer rest wavelengths shift into the $I$-band, such objects
quickly fade to the point of invisibility.  We have computed synthetic
$I$ magnitudes as a function of redshift for our sample spectra to
determine the limit of visibility ($I<20.5$) for each source.  Distances were
computed using a $\Lambda$-cosmology with $H_0=70$, $\Omega_m = 0.3$,
and $\Omega_\Lambda = 0.7$.  The results for the ten $z<1.3$ quasars
are given in Table~3.  If the red quasars that we observe were moved to
$z=1.5$, none of them would have been detectable above the $I<20.5$
limit of our survey.  In contrast, the low-redshift blue quasars
detected in our sample are visible to much greater redshifts, typically
$z=1.7$--2.1.  Unless red quasars are preferentially found only at low
redshifts --- and it is just a matter of luck that our detection limit
forbids us from detecting more distant objects --- there must be many
more red quasars associated with the quasar population peak at
$1<z<2$.

The excellent fits to the spectra in Figure~\ref{fig-reddening}(b)
argue strongly in favor of dust as the source of the reddening for
at least some of our red quasars.  In that case the implied
dereddened luminosities are large.
Even with a magnitude
limit $I<20.5$, our survey is deep enough to detect only the most
luminous of these reddened quasars. Figure~\ref{fig-absmag} shows the
dereddened $I$-band absolute magnitudes as a function of redshift along
with the effect of reddening on our detection limit.
For higher redshift objects where bluer rest wavelengths move into the
$I$-band, even a
small amount of dust absorption has a dramatic effect on
detectability.  The only objects detectable at a reddening as large as
$E(B-V)=0.7$ are the three quasars that are actually seen to be very
red (Fig.~\ref{fig-reddening}b),
and it would be impossible to detect similar objects in the
$I$-band at redshifts greater than $\sim1.5$.
The five most luminous
objects with $z<1.3$ are {\it all} red.

Unless highly reddened quasars are also preferentially highly luminous,
there must be an even larger, as yet undiscovered, population of
quasars at lower luminosity.  Such a correlation is not completely
implausible; perhaps quasars that are heavily reddened have just
accreted a new source of fuel that guarantees they will also be
relatively luminous.  Since the luminosity cannot exceed the Eddington
limit, however, we still expect to find lower luminosity
reddened quasars associated with lower black hole masses.

This conclusion appears difficult to escape.  For some quasars, an
optical synchroton emission component or contaminating light from a
host galaxy can produce a redder spectrum (Benn et al.\ 1998; Whiting,
Webster \& Francis 2001).  Our red quasars with lower $E(B-V)$ values
might be fitted with a red synchrotron continuum.  However, any
explanation of the reddest quasars based on an additional emission
mechanism is strongly ruled out by our observations.  Not only are the
reddened quasar template spectra generally a good fit to our data, but
the quasar light must be strongly suppressed in the ultraviolet to
explain the absence of red quasars at higher redshifts from our
sample.

Our data strongly support the hypothesis, originally raised by Webster
et al.\ (1995) based on a radio sample with a much higher
($\sim100$~mJy) flux density threshold, that radio quasars are
dominated by a previously undetected population of red, heavily
obscured objects (see also Srianand \& Kembhavi 1997; Benn et
al.\ 1998; Kim \& Elvis 1999).  The red objects we detect represent fully
half the population of the $z<1.3$ quasars in our sample; unless their
underlying luminosity and redshift distributions are {\it very}
different than those of blue quasars, there must be many more such
objects at higher redshifts and lower luminosities comparable to our
blue quasars.  We are likely to be detecting only the most luminous tip
of the red quasar iceberg.

\section{Associations Between Quasars and Clusters}

We have compared the positions of the objects in our sample to the
catalog of cluster candidates in Postman et al.\ (2002). Eight sources
lie within $2.5^{\prime}$ of a cluster centroid, with an expected
chance coincidence rate of five matches (ignoring the small overlap of
cluster candidates).  Five of the cluster/radio-source
coincidences are matches between rarer high-redshift cluster candidates and
confirmed quasars. A $2.5^{\prime}$ radius around the $z\ge 0.8$
clusters covers only 3.7\% of the survey area, so that only 1.3 chance
coincidences with quasars would be expected. In addition, three of the
matching quasars have redshifts similar to the estimated cluster
redshifts: cluster \#243 ($z\sim1.1$) with FDQ~J101510.6+523643
($z=0.960$), cluster \#304 ($z\sim0.8$) with FDQ~J101119.2+520536
($z=0.901$), and cluster \#382 ($z\sim0.8$) with FDQ~J101002.1+512700
($z=0.787$). It is likely that all three are real associations that
confirm the cluster redshift estimates.

Two of the three quasars
coincident with X-ray sources are also close to cluster
candidates, and it is possible that a hot intracluster medium is
responsible for the X-ray emission; in the case of cluster \#34
($z\sim0.3$) and FDQ~J102256.6+540717 ($z=0.340$), the cluster and
quasar redshifts match, and a physical association is quite probable.

There are only two other FDQ quasars with redshifts less than unity, and
one of them
also has a cluster match, albeit at a slightly greater separation.
Cluster \#303 ($z\sim0.7$) is $3.0^\prime$ from FDQ~J102059.8+520917
($z=0.813$).  The close agreement in the redshifts makes this another
likely physical association.

This is a remarkable result: five of the six $z<1$ quasars in the FDQ
are associated with Deeprange cluster candidates of similar estimated
redshifts.  Despite the large number of candidates (444) in the cluster
catalog, this is highly statistically significant.  Using a radius of
$3^\prime$, the cluster candidates cover 22\% of the survey area, so
the probability that five of six $z<1$ quasars lie within $3^\prime$ of a
cluster of {\em any} redshift is only $2\times10^{-3}$.  The
probability that five randomly selected clusters will all have redshifts
within $\pm0.15$ of the five $z<1$ quasar redshifts is even smaller,
only $\sim5\times10^{-4}$.  The joint probability of both happening by
chance is very small, even considering the {\it a posteriori} nature of this
argument.

The only $z<1$ quasar without an
associated cluster is FDQ~J102132.2+511433 ($z=0.944$). The
Deeprange cluster selection efficiency is expected to decline
at such high redshifts (Postman et al.\ 2002), so we would
not expect all $z\sim0.95$ clusters to be included in the
Deeprange catalog.  Our results are therefore consistent with the
possibility that all of our $z<1$ quasars are associated with clusters.

This has some surprising implications.  Recall that only 10\% of quasars
are radio-loud (RLQs); there must be $\sim300$ additional radio-quiet
QSOs (RQQs) with $I<20.5$ in the Deeprange area.  Since 3 of the 109
$z\ge0.8$ Deeprange cluster candidates have associated RLQs, one can
project that $\sim30$ such clusters will have an associated quasar when the
RQQs are included.  Since 30--40\% of the $z\ge0.7$ cluster candidates
are expected to be spurious (Postman et al.\ 2002), that would lead to
half of the real high-redshift cluster candidates having associated
quasars.

Another possibility, however, is that the RLQs are preferentially found
in clusters and that the (as yet undiscovered) RQQs in this field will
not be associated with Deeprange clusters.  Many studies have concluded
that RLQs are more likely to be found in high-density cluster
environments than are RQQs (e.g., Yee \& Green 1984; Ellingson, Yee \&
Green 1991) although others have concluded that RQQs and RLQs
are both found in clusters (e.g., Hutchings, Crampton \& Johnson 1995;
Wold et al.\ 2001).  Recently S{\" o}chting, Clowes, \& Campusano
(2002) explored the correlation between clusters and RQQs at $z\sim0.3$
and found that the RQQs do associate with clusters but that they tend
to occupy the cluster periphery rather than the cluster core.  Note
that some of the five FDQ quasars are also relatively far from the cluster
centers, with projected local distances of 0.4--1.5~Mpc.

Quasar/cluster associations have generally not been analyzed
at $z\sim1$ because it is difficult to identify clusters at such
high redshifts.  Since quasars were far more common at $z\sim1$ than at
$z\sim0.3$, it is expected that more clusters will have quasars at high
redshifts.  Whether as many as half the $z\sim1$ clusters have
associated quasars, as we suggest, can be confirmed by future studies.
The Deeprange field would be a fruitful target for searches for
radio-quiet quasars in order to quantify this association.  Spectroscopic
redshifts for the associated cluster candidates will also help confirm
their association with FDQ quasars.

\section{Summary and Conclusions}

A complete, magnitude-limited sample of radio-emitting quasars in the
Deeprange I-band survey has been constructed by obtaining optical spectra
for 50 stellar counterparts to {\it FIRST} radio sources in the 16 deg$^2$
survey region. Thirty-five quasars were detected. In addition to the
first radio-loud FRII BAL quasar, two additional BAL quasars were
discovered, representing fully one-third of objects with a high enough redshift
for the BAL signature to be seen. A probable damped-Lyman-alpha absorber was
identified in another source. Most intriguing, however, is the discovery
of five very red quasars with $0.7<z<1.3$ that represent half
of all objects in this redshift range.
Indeed, if we use dereddened magnitudes,
the five most luminous
quasars with $z<1.3$ are {\it all} red, and three of the six most luminous
objects in the entire 35-member sample are red.
Fits of the FBQS
composite quasar spectrum with an SMC reddening law yield $I$-band extinctions
of $1.2<A(I)<3.7$ magnitudes for the five reddest quasars.

We show that with reddenings this large, even the most luminous quasars
($M_B<-27$) are not detectable beyond $z=1.5$ at our $I$-band magnitude limit
of 20.5. Indeed, the combination of the good fits of the reddened composite
spectrum to our spectra and the absence of any red quasars at higher redshift
(naturally attributable to extinction) strongly suggest that 1) the quasars
are red because of dust, and 2) the detected objects represent only a small
fraction of the total population.

This striking support for the Webster et al.\ (1995) results on the
missing red quasar population, using the much larger population of
objects with radio fluxes several hundred times fainter, is cause for
reflection.  Are the most luminous quasars red because they are both
magnified and reddened by an intervening galaxy?  Gregg et al.\ (2002)
argued that this could explain the red quasars found in a survey
of {\it FIRST}-2MASS counterparts, and several other examples are known
of very red quasars that are lensed (Hewitt et al.\ 1992, Courbin et al.\
1998).
An argument that weighs
heavily against this hypothesis is that the lensing fraction would be
absurdly high.  If most lensed quasars are missed because they
are reddened, it is in principle possible that we have the lensing
fraction wrong; however, cosmological constraints strongly reject the
possibility that the lensing fraction is as high as 10\%.
A high lensing fraction in objects selected to be very red is
allowed since that sample could be strongly biased, but the
sample presented in this paper has no color bias (for either red or
blue objects) so it should not have any particular bias toward
lensed objects.

Could the dust fraction increase with cosmic time (along with
metallicity) so that moving out beyond $z=1.3$ will not yield a
significant additional population of reddened quasars?  Possibly,
although there is little evidence from quasar spectra for metallicity
evolution out to $z>6$; it is more likely that dust gets
destroyed and/or blown away as a quasar ages and undergoes repeated
high luminosity intervals, so that there might well be fewer reddened
objects at low redshift.

It remains unknown whether or not the large number of hidden quasars
adduced from first the radio-loud and now the radio-intermediate segments
of the population extends to radio-quiet objects. We note that
the radio flux distribution for the red quasars is indistinguishable
from that for the other quasars in our sample, and that two of the
red quasars have flux densities close to the {\it FIRST} detection
limit.  It is therefore unlikely that such objects are found only
in radio sources with $F_\nu>1$~mJy.
Addressing the question of whether red quasars are common in
the radio-quiet population
is clearly of considerable importance for assessing the true accretion
luminosity of the Universe and its evolution with cosmic time.

Finally, we have found that five of the six quasars in our sample with $z<1$
are associated with Deeprange cluster candidates with concordant redshift
estimates. Extrapolating these results to the radio-quiet portion of the
population suggests that up to half of the $0.3<z<1$ clusters have
associated quasars. Spectroscopic redshifts for the cluster galaxies and
a search for radio-quiet quasars in this field are required to confirm
this suggestion.

\acknowledgments

We thank the referee, Michael Drinkwater, for some very helpful
comments which led us to quantify and strengthen our conclusions about
the implications of these results for the missing red quasar population.
The success of the {\it FIRST} survey is in large measure due to the
generous support of a number of organizations. In particular, we
acknowledge support from the NRAO, the NSF (grants AST-00-98259 and
AST-00-98355), the Institute of Geophysics and
Planetary Physics (operated under the auspices of the U.~S.\ Department
of Energy by Lawrence Livermore National Laboratory under contract No.
W-7405-Eng-48), the Space Telescope Science Institute, NATO, the
National Geographic Society (grant NGS No.~5393-094), Columbia
University, and Sun Microsystems.  DJH is grateful for the support of
the Raymond and Beverly Sackler Fund, and joins RHB and RLW in thanking
the Institute of Astronomy of the University of Cambridge for
hospitality during some of this work.

\clearpage

\thispagestyle{empty}

\begin{deluxetable}{ccrrccccrrrrll}
\rotate
\tabletypesize{\scriptsize}
\tablecolumns{14}
\tablewidth{0pc}
\tablecaption{Spectroscopically Confirmed Quasars in the Deeprange Field}
\tablehead{
\colhead{RA} & \colhead{Dec} & \colhead{Offset} & \colhead{$F_{pcore}$} & \colhead{Size} & \colhead{$F_{int}$} & \colhead{NVSS} & \colhead{$I$} & \colhead{$F$\tablenotemark{a}} & \colhead{$J$\tablenotemark{a}} & \colhead{$E(B{-}V)$\tablenotemark{b}} & \colhead{$A(I)$\tablenotemark{b}} & \colhead{$z$} & \colhead{Comments\tablenotemark{c}}\\
\colhead{(2000)} & \colhead{(2000)} & \colhead{($^{\prime\prime}$)} & \colhead{(mJy)} & \colhead{($^{\prime\prime}$)} & \colhead{(mJy)} & \colhead{(mJy)}
& \colhead{(mag)} & \colhead{(mag)} & \colhead{(mag)} & \colhead{(mag)} & \colhead{(mag)} \\
\colhead{(1)} & \colhead{(2)} & \colhead{(3)} & \colhead{(4)} & \colhead{(5)} & \colhead{(6)} & \colhead{(7)} &
\colhead{(8)} & \colhead{(9)} & \colhead{(10)} & \colhead{(11)} & \colhead{(12)} & \colhead{(13)} & \colhead{(14)}
}
\startdata
10 01 30.555 &  +53 31 51.02 &    0.49 &  4.0 &  $3.1\times 2.5$ &   5.1 &   $4.2\pm 0.4$ &  19.74 &   $>20.4$ &   20.84 & $-0.04$ & $-0.29$ & 2.192 & \\
10 01 32.427 &  +51 29 54.52 &    2.06 &  1.4 &  $4.4\times 0.0$ &   1.8 &  $<1.4$ &       19.06 &  19.84 &  20.54 & $-0.05$ & $-0.24$ & 1.094 & \\
10 02 40.947 &  +51 12 45.36 &    0.83 &  3.8 &   $<2.5$ &     \nodata &   $4.0\pm 0.4$ &  20.34 &   $>20.4$ &   21.44\tablenotemark{*} & $-0.04$ & $-0.24$ & 1.80 &  close pair; poss.~damped Ly$\alpha$, $z=0.99$ Mg II\\
10 03 11.535 &  +50 57 05.39 &    1.44 &  1.6 &   $\sim 55T$ &    24.3 &  $24.2\pm 1.2$ &  19.33 &  18.98 &  19.93 & $0.01$ & $ 0.03$ & 1.835 & close pair; poss.~damped Ly$\alpha$, $z=0.91$ Mg II\\
10 03 50.707 &  +52 53 52.41 &   13.08 &  6.9 &   $\sim 55T$ &   113.0 & $114.8\pm 3.8$ &  18.52 &  18.58 &  19.59 & $-0.02$ & $-0.09$ & 1.33 &  \\
10 03 53.654 &  +51 54 57.29 &    0.50 &  1.1 &  $6.3\times 0.0$ &   1.5 &  $<1.4$ &       18.26 &  18.85 &  19.88 & $0.12$ & $ 0.70$ & 1.63 &  v.~broad Mg II; strong Fe II\\
10 04 00.401 &  +54 16 16.96 &    0.83 &  9.0 &   $<2.5$ &     \nodata &   $7.1\pm 0.5$ &  18.84 &  19.17 &  20.52 & $0.06$ & $ 0.31$ & 1.48 &  \\
10 06 32.372 &  +53 48 52.08 &    0.34 &  1.3 &  $3.0\times 0.0$ &   1.4 &   $4.8\pm 0.5$ &  19.22 &  20.40\tablenotemark{*} & 22.15 & $0.82$ & $ 3.61$ & 1.040 & very red; confused \&/or extended ($F_{int}=3.9$~mJy)\\
10 08 08.472 &  +50 43 21.36 &    0.17 & 17.0 &   $<2.5$ &     \nodata &  $17.3\pm 0.6$ &  19.57 &  19.50 &  20.06 & $0.03$ & $ 0.14$ & 1.43 &  \\
10 09 04.530 &  +52 44 17.37 &    2.03 &  1.9 & $13.8\times 4.3$ &   8.7 &   $7.6\pm 0.5$ &  19.79 &  19.58 &  20.70 & $-0.02$ & $-0.10$ & 1.45 &  \\
10 10 02.117 &  +51 27 00.20 &    1.21 &  1.4 &  $3.7\times 1.1$ &   1.8 &  $\sim 2.0$ &       18.01 &  18.38 &  18.44 & $-0.07$ & $-0.29$ & 0.787 & \\
10 10 39.305 &  +51 31 09.04 &    0.64 & 11.3 &   $<2.5$ &     \nodata &  $12.3\pm 0.5$ &  20.00 &  19.81 &  20.82 & $-0.03$ & $-0.24$ & 2.562 & \\
10 11 19.296 &  +52 05 36.54 &    0.72 &  1.2 &   $<2.5$ &     \nodata &   $3.1\pm 0.5$ &  20.00 &   $>20.4$ &    $>22.4$ & $0.90$ & $ 3.73$ & 0.901 & very red\\
10 12 04.077 &  +53 13 31.91 &    0.36 &  3.5 &   $<2.5$ &     \nodata &   $7.0\pm 1.4$ &  18.54 &  18.88 &  19.57 & $0.02$ & $ 0.15$ & 2.911 & $\sim 100^{\prime\prime}$ extent in NVSS; very weak lines\\
10 13 08.804 &  +50 50 17.68 &    0.45 &  9.0 &   $<2.5$ &     \nodata &  $13.4\pm 0.9$ &  19.34 &  19.78 &  20.15 & $-0.03$ & $-0.21$ & 1.817 & $\sim 26^{\prime\prime}$ extent in NVSS; Mg II abs\\
10 13 16.827 &  +51 11 18.06 &    0.77 & 18.4 &   $<2.5$ &     \nodata &  $16.0\pm 0.6$ &  19.87 &   $>20.4$ &    $>22.4$ & $0.26$ & $ 1.20$ & 1.16 &  red; low S/N spectrum\\
10 14 09.896 &  +51 52 28.67 &    1.20 &  3.1 &   $<2.5$ &     \nodata &  $\sim 2.0$ &       19.54 &  19.79 &  20.36 & $0.01$ & $ 0.06$ & 1.775 & \\
10 15 10.623 &  +52 36 43.22 &    0.20 &  4.7 &   $3.7\times 1.2$ &  5.8 &   $6.8\pm 0.5$ &  20.15 &  20.45 &  22.04 & $0.10$ & $ 0.44$ & 0.960 & \\
10 16 14.223 &  +52 09 15.68 &    0.30 &  5.2 &    $\sim 62T$ &  176.9 & $185.2\pm 6.0$ &  18.68 &  19.19 &  20.92 & \nodata & \nodata & 2.43 &  FRII HiBAL; weak emission lines\\
10 16 15.842 &  +50 55 13.65 &    0.29 & 27.8 &   $<2.5$ &     \nodata &  $26.8\pm 0.9$ &  20.25 &  20.39 &  21.07 & $0.01$ & $ 0.12$ & 3.14 &   strong rest C IV abs\\
10 17 35.005 &  +53 35 35.49 &    0.44 &  1.8 &  $2.8\times 0.0$ &   1.9 &  $\sim 2.0$ &       19.39 &   $>20.4$ &    $>22.4$ & \nodata & \nodata & 3.27 &  bright foreground cluster; spectacular LoBAL\\
10 17 42.660 &  +53 56 35.50 &    0.76 & 67.8 &  $3.8\times 1.2$ &  85.2 &  $92.9\pm 2.8$ &  18.43 &  18.67 &  19.72 & $0.05$ & $ 0.29$ & 1.40 &  poss.~damped Ly$\alpha$, $z=1.305$ Mg II\\
10 18 57.724 &  +53 34 13.49 &    0.32 &  8.9 &   $<2.5$ &     \nodata &   $7.9\pm 0.5$ &  19.59 &  19.83 &  20.78 & $-0.03$ & $-0.16$ & 1.495 & \\
10 19 16.385 &  +50 45 57.50 &    0.30 &  1.9 &  $4.8\times 3.2$ &   2.9 &   $4.5\pm 0.5$ &  17.37 &  17.29 &  18.07 & $0.00$ & $ 0.01$ & 1.311 & \\
10 19 39.990 &  +51 49 24.04 &    0.51 &  3.6 &   $<2.5$ &     \nodata &   $3.8\pm 0.4$ &  20.54 &  19.93\tablenotemark{*} & 20.19\tablenotemark{*} & $0.07$ & $ 0.38$ & 1.502 & note $I>20.5$; faint continuum \\
10 20 07.209 &  +52 24 45.98 &    0.96 &  6.5 &  $2.7\times 1.3$ &   7.5 &   $6.8\pm 0.5$ &  19.74 &  19.89\tablenotemark{*} &  $>22.4$ & $0.56$ & $ 2.73$ & 1.225 & very red \\
10 20 50.635 &  +51 57 19.01 &    0.52 &  1.9 &   $<2.5$ &     \nodata &   $4.4\pm 0.5$ &  19.41 &  19.65\tablenotemark{*} & 20.49 & $0.06$ & $ 0.34$ & 1.63 &  unrelated source $12^{\prime\prime}$ away\\
10 20 59.887 &  +52 09 17.87 &    0.60 &  1.4 &   $\sim 120T$ &   35.1 & $34.0\pm 1.5$ &  18.67 &  18.88 & 19.63 & $0.01$ & $ 0.02$ & 0.813 & \\
10 21 01.454 &  +52 39 56.94 &    0.22 &  3.8 &   $<2.5$ &     \nodata &  $<1.4$ &       18.65 &  18.58 &  19.39 & $-0.03$ & $-0.14$ & 1.492 & PSPC=0.010\\
10 21 32.233 &  +51 14 33.80 &    0.24 & 63.7 & $3.2\times1.0$ &  75.1 &    87.0 &       19.91 & $>20.4$ &  21.44 & $0.40$ & $ 1.60$ & 0.944 & red; core-jet source; PSPC = 0.008 \\
10 22 23.869 &  +52 38 42.50 &    0.60 &  6.3 &   $<2.5$ &     \nodata &   $7.2\pm 0.4$ &  18.83 &  19.48 &  20.71 & $0.01$ & $ 0.05$ & 1.857 & HiBAL\\
10 22 56.641 &  +54 07 17.95 &    0.41 &  1.2 &   $<2.5$ &     \nodata &  $\sim 1.4$ &       18.09 &  17.97 &  19.25 & $0.04$ & $ 0.11$ & 0.340 & RASS=0.024\\
10 23 52.706 &  +54 06 49.63 &    0.77 &  1.6 &   $<2.5$ &     \nodata &  $<1.4$ &       19.96 &  20.30 &  21.22 & $-0.04$ & $-0.24$ & 1.95 &  \\
10 24 22.390 &  +51 38 38.77 &    0.78 &  2.2 &  $6.6\times 3.5$ &   4.1 &   $5.1\pm 0.4$ &  19.42 &  20.24 &  20.80 & $-0.04$ & $-0.24$ & 1.518 & 6.15mJy with 2nd source $25^{\prime\prime}$ away\\
10 26 27.697 &  +51 41 14.49 &    0.53 &  5.5 &   $<2.5$ &     \nodata &   $6.6\pm 0.4$ &  18.54 &  19.23 &  19.71 & $-0.01$ & $-0.10$ & 2.63 &  27mJy @6cm $\rightarrow$ inverted spectrum\\
\tablenotetext{a}{$J$ and $F$ magnitudes (roughly $B$ and $R$) are from
the GSC-2 catalog of the POSS-II photographic plates. Magnitudes marked
with an asterisk (*) are from the APM POSS~I catalog, and have been
calibrated as described in McMahon et al.\ (2002);\\ magnitude limits
for the APM are $E<20.1$ and $O<21.95$ and the magnitude limits for
GSC-2 are $F<21.4$ and $J<22.4$.}
\tablenotetext{b}{Dust reddening was derived from model fits as described in the text.
$E(B-V)$ is the rest frame reddening, while $A(I)$ is the extinction in the
observed $I$-band.}
\tablenotetext{c}{RASS = Rosat All-Sky Survey (Voges et al.\ 1999) and
PSPC = serendipitous detection in a pointed ROSAT observation.}
\enddata
\end{deluxetable}

\clearpage

\begin{deluxetable}{ccccccccrl}
\tabletypesize{\scriptsize}
\tablecolumns{10}
\tablewidth{0pc}
\tablecaption{FIRST Deeprange Sample Non-Quasars}
\tablehead{
\colhead{RA} & \colhead{Dec} & \colhead{Offset} & \colhead{$F_{\rm pcore}$} & \colhead{Size} & \colhead{$F_{\rm int}$} & \colhead{NVSS} & \colhead{$I$} &
\colhead{$z$} & \colhead{Comments\tablenotemark{a}} \\
\colhead{(2000)} & \colhead{(2000)} & \colhead{(${}^{\prime\prime}$)} & \colhead{(mJy)} & \colhead{(${}^{\prime\prime}$)} & \colhead{(mJy)} & \colhead{(mJy)} & \colhead{(mag)} &
\colhead{} & \colhead{} \\
\colhead{(1)} & \colhead{(2)} & \colhead{(3)} & \colhead{(4)} & \colhead{(5)} & \colhead{(6)} & \colhead{(7)} &
\colhead{(8)} & \colhead{(9)} & \colhead{(10)}
}
\startdata
10 00 40.735 &  +53 19 11.69 & 0.27 &     2.7 &  $<2.5$ &       \nodata &      3.0 & 18.77 & \nodata &  BL Lac object \\
10 01 34.174 &  +50 43 37.98 & 1.46 &     2.6 &  $4.4\times 0.0$ &  3.0 &      3.8 & 18.61 &  0.39 &  Galaxy \\
10 01 52.743 &  +52 46 13.43 & 0.48 &     1.1 &  $<2.5$ &       \nodata &     $<1$ & 20.09 &  0.56 &  H II Galaxy \\
10 02 53.517 &  +51 37 08.56 & 1.20 &     1.4 &  $<2.5$ &       \nodata & $\sim 1$ & 17.08 &  0.25 &  H II Galaxy \\
10 03 25.102 &  +51 01 18.76 & 0.37 &     1.6 &  $<2.5$ &       \nodata & $\sim 1$ & 20.26 &  0.66 &  H II Galaxy (surrounding cluster) \\
10 04 27.868 &  +54 11 55.93 & 1.07 &     1.8 &  $6.6\times 4.2$ &  3.7 &      2.3 & 19.55 &  0.45 &  H II Galaxy \\
10 05 12.387 &  +54 32 40.99 & 0.32 &     1.4 &  $>400$ &          40.2 &  121-177 & 19.27 &  0.49 &  H II Galaxy \\
10 11 45.770 &  +51 28 49.82 & 1.07 &     1.2 &  $7.3\times 1.4$ &  2.2 & $\sim 1$ & 19.77 &  0.69 &  Galaxy \\
10 12 33.446 &  +53 07 02.08 & 0.53 &     1.7 &  $<2.5$ &       \nodata &      4.2 & 19.35 &  0.00 &  Star \\
10 14 43.644 &  +51 23 44.34 & 0.10 &     2.6 &  $<2.5$ &       \nodata &      2.5 & 17.08 &  0.30 &  Galaxy \\
10 16 36.503 &  +52 52 37.76 & 5.94 & \nodata &  $\sim 15$ &      10.0D &     10.1 & 19.99 & $\sim 0.90$ &  Galaxy \\
10 23 08.712 &  +52 47 40.18 & 0.64 &     1.6 &  $4.5\times 0.0$ &  2.0 & $\sim 1$ & 20.36 &  0.93 &  H II Galaxy \\
10 23 43.782 &  +51 23 46.90 & 3.85 & \nodata &  $\sim 10$ &       21.3 &     21.9 & 19.56 &  0.51 &  Galaxy \\
10 25 39.038 &  +51 30 25.15 & 1.05 &     1.1 &  $<2.5$ &       \nodata &     $<1$ & 19.97 &  0.75 &  AGN \\
10 26 07.741 &  +52 20 39.65 & 1.49 &     2.2 &  $5.3\times 2.2$ &  3.4 & $\sim 1$ & 19.43 &  0.44 &  AGN \\
\tablenotetext{a}{Classification criteria are taken from FBQS2 and FBQS3 and are discussed
in the text.}
\enddata
\end{deluxetable}

\begin{deluxetable}{cccccc}
\tablecolumns{6}
\tablewidth{0pc}
\tablecaption{Redshift Detection Limits for $z<1.3$ FDQ Quasars}
\tablehead{
\colhead{RA} & \colhead{Dec} & \colhead{$I$} & \colhead{$E(B{-}V)$} & \colhead{$z$} & \colhead{$z_{max}$} \\
\colhead{(2000)} & \colhead{(2000)} & \colhead{(mag)} & \colhead{(mag)} \\
\colhead{(1)} & \colhead{(2)} & \colhead{(3)} & \colhead{(4)} & \colhead{(5)} & \colhead{(6)}
}
\startdata
10 22 56.641 & +54 07 17.95 & 18.09 & \phs 0.04  & 0.340 &  0.85 \\
10 10 02.117 & +51 27 00.20 & 18.01 &   $-$0.07  & 0.787 &  2.12 \\
10 20 59.887 & +52 09 17.87 & 18.67 & \phs 0.01  & 0.813 &  1.70 \\
10 11 19.296 & +52 05 36.54 & 20.00 & \phs 0.90  & 0.901 &  1.01 \\
10 21 32.233 & +51 14 33.80 & 19.91 & \phs 0.40  & 0.944 &  1.09 \\
10 15 10.623 & +52 36 43.22 & 20.15 & \phs 0.10  & 0.960 &  1.06 \\
10 06 32.372 & +53 48 52.08 & 19.22 & \phs 0.82  & 1.040 &  1.43 \\
10 01 32.427 & +51 29 54.52 & 19.06 &   $-$0.05  & 1.094 &  2.12 \\
10 13 16.827 & +51 11 18.06 & 19.87 & \phs 0.26  & 1.160 &  1.42 \\
10 20 07.209 & +52 24 45.98 & 19.74 & \phs 0.56  & 1.225 &  1.51 \\
\enddata
\end{deluxetable}

\end{document}